\documentclass[preprint,12pt]{elsarticle}
\usepackage{aas_macros}
\usepackage{amssymb}
\usepackage{amsmath}
\usepackage[dvipsnames]{xcolor}
\usepackage{enumitem}
\usepackage[colorlinks=true,
            linkcolor=MidnightBlue,  
            citecolor=ForestGreen,   
            urlcolor=BrickRed]       
           {hyperref}
\journal{Astronomy & Computing}
\biboptions{sort&compress}
\usepackage{subcaption}
\usepackage{tabularx}
\usepackage{array}
\usepackage{booktabs}
\usepackage{comment}
\usepackage{multirow}
\usepackage{graphicx}

\usepackage{etoolbox}

\AtBeginEnvironment{thebibliography}{%
  \small                      
  \setlength{\itemsep}{0pt}   
  \setlength{\parskip}{0pt}   
  \setlength{\bibsep}{0pt plus 0.3ex}
}

\begin{document}

\begin{frontmatter}

\title{A Practical Study of Lightweight Neural Gravitational-Wave Detection in Real LIGO Noise: Models, Ablations, and Open Software}

\author[1,2,3]{Victoria Tiki\corref{cor1}}
\ead{vtiki2@illinois.edu}

\author[1,2,3,5,6]{E.~A. Huerta}

\cortext[cor1]{Corresponding author}

\affiliation[1]{organization={Data Science and Learning Division, Argonne National Laboratory},
            city={Lemont},
            state={Illinois},
            postcode={60439},
            country={USA}}

\affiliation[2]{organization={Department of Physics, University of Illinois Urbana-Champaign},
            city={Urbana},
            state={Illinois},
            postcode={61801},
            country={USA}}

\affiliation[3]{organization={National Center for Supercomputing Applications, University of Illinois Urbana-Champaign},
            city={Urbana},
            state={Illinois},
            postcode={61801},
            country={USA}}

\affiliation[5]{organization={Department of Computer Science, The University of Chicago},
            city={Chicago},
            state={Illinois},
            postcode={60637},
            country={USA}}

\affiliation[6]{organization={Department of Astronomy, University of Illinois Urbana-Champaign},
            city={Urbana},
            state={Illinois},
            postcode={61801},
            country={USA}}

\begin{abstract}
Machine-learning methods offer a computationally efficient complement to conventional gravitational-wave searches, but their performance depends strongly on training data, preprocessing, inference design, and robustness to nonstationary detector noise. We present an open-source framework for developing and evaluating lightweight neural detectors using real LIGO data. The pipeline includes strain acquisition, PSD estimation, waveform injection, training, inference, and trigger construction. Using O3a data, we compare twelve architectures and perform ablations of normalization, whitening, label width, glitch handling, SNR curricula, inference triggers, seeds, and signal populations. We find that simple temporal convolutional networks match or outperform more complex architectures, including attention-based and graph models. Shared normalization improves stability, intermediate whitening is optimal, and including sub-threshold signals improves weak-signal recovery. We then train four independently initialized 116k-parameter TCNs on O3b noise and evaluate them on 22.7 days of coincident data containing 13 confident GWTC-3 events. At a threshold fixed using O3a data, the four models recover 8, 9, 9, and 9 events while producing 0, 0, 1, and 0 false-positive triggers, respectively, without post hoc vetoes or inter-model coincidence filtering. In injection studies, the models achieve a median sensitive volume of \(5.4\,\mathrm{Gpc}^3\) at the preselected threshold, corresponding to a median false-alarm rate of \(20.5\,\mathrm{yr}^{-1}\). At a stricter threshold, the median sensitive volume remains \(4.6\,\mathrm{Gpc}^3\), with three models returning 0 false alarms and one model yielding 1 false alarm, observed across one year of background per model. These results show that, within the tested implementations and training setup, careful pipeline design can matter more than architectural complexity and provide a lightweight pipeline for future studies on real interferometer data.\end{abstract}


%

\end{frontmatter}


\section{Introduction}
\label{sec:intro}

Gravitational-wave (GW) observations provide a direct probe of compact-object populations and strong-field gravity~\cite{2019ApJ...882L..24A,Zevin_2020,PhysRevLett.123.011102,PhysRevLett.123.111102}. Detecting these signals requires identifying short, weak transients from binary black hole (BBH), binary neutron star (BNS), and neutron star--black hole (NSBH) mergers in nonstationary interferometer noise. Rapid identification is additionally important for events with electromagnetic counterparts, for which low-latency alerts enable prompt follow-up observations~\cite{LIGOScientific:2019gag}.

Matched-filter pipelines such as PyCBC~\cite{Usman:2015kfa} and GstLAL~\cite{Messick_2017} remain the standard approach to compact-binary searches. These methods correlate detector strain against waveform-template banks and provide the statistical framework used for current production analyses. Their success has enabled the detection of increasingly large compact-binary populations, but the computational cost of template-based searches and the complexity of distinguishing signals from transient detector artifacts motivate the development of complementary detection methods~\cite{Verma_2022,Shawhan_2004,PhysRevD.60.022002}.

Machine-learning methods offer one such complement. Early studies showed that CNNs could approach matched-filter sensitivity for simulated BBH signals in Gaussian noise~\cite{George_2018,Gabbard2018}. Gebhard et al.\ subsequently evaluated a fully convolutional network as a trigger generator using real LIGO noise and emphasized the limitations of standard classification metrics for search-level evaluation~\cite{Gebhard2019}. Wang et al.\ trained CNNs on O1 noise and recovered the cataloged BBH events in O1 and O2 observing runs~\cite{PhysRevD.101.104003}, while Qiu et al.\ extended neural classification to BBH, BNS, and NSBH signals injected into real detector noise~\cite{QIU2023137850}. Tian et al.\ subsequently applied an ensemble of hierarchical convolutional and graph-based models to February 2020 data~\cite{Tian:2023vdc}. More recent systems have moved toward more complete search pipelines: Aframe uses a ResNet-based detection statistic for low-latency compact-binary searches~\cite{Marx2025Aframe}, while AresGW applies a deep ResNet to an extended O3 search and incorporates candidate classification and statistical interpretation~\cite{Nousi_2023,Koloniari_2025}. Machine learning has also been used to rescore triggers produced by conventional matched-filter pipelines rather than replace the initial search stage~\cite{mobilia2025machinelearningassessastrophysical}.

The performance of a neural GW detector is not determined by its architecture alone. Real interferometer data contains detector-dependent transient artifacts, making noise selection and glitch handling important components of the detection pipeline~\cite{Glanzer_2023,LIGO:2021ppb,Zevin_2017}. It also depends on how real noise is selected and cleaned, how strain is whitened and normalized, which signal population and signal-to-noise-ratio (SNR) distribution are used during training, how targets are defined, which checkpoint and training seed are evaluated, and how continuous model outputs are converted into candidate triggers. Differences in these choices make results across architectures and studies difficult to interpret. Real-noise evaluations may additionally incorporate coincidence among several independently trained networks, external glitch catalogs, data-quality vetoes, or manual candidate review. Such stages can be appropriate components of a complete search, but they make it harder to isolate the robustness of the learned detection statistic itself. A systematic study of the full neural detection pipeline under common data and evaluation conditions is therefore needed.

In this work, we present an open-source framework for developing and evaluating lightweight neural GW detectors using real LIGO data. The software covers the full workflow from GWOSC strain acquisition and power-spectral-density estimation through waveform injection, training, checkpoint selection, continuous inference, trigger construction, and sensitivity evaluation. We use O3a data as a development set to compare twelve neural architectures and to study variations in normalization, whitening context, target-label width, glitch handling, SNR curriculum, inference-trigger construction, training seed, and signal population. The architectures include temporal convolutional networks (TCNs), hierarchical dilated convolutional networks, attention- and graph-based models, a ResNet, a time--frequency model, and a state-space model. We evaluate all retained models within a common pipeline using sensitive volume as a function of the observed background count. We also introduce an amplitude-spectral-density mismatch measure that partitions evaluation noise according to its similarity to the training distribution, providing a model-independent diagnostic of noise-domain shift.

The primary contributions of this work are therefore: (i) an open, end-to-end software framework for neural GW detection using real GWOSC strain; (ii) a controlled comparison of model architectures together with broad ablations of the training, preprocessing, and inference pipeline; (iii) a noise-similarity diagnostic and evaluation protocol designed to expose seed dependence and distribution shift; and (iv) a lightweight single-model configuration validated on a held-out O3b interval with minimal trigger processing. The public repository~\cite{AttenGWRepo} retains the name
\textsc{AttenGW} for historical reasons: it was originally developed for a cross-attention model introduced as part of the RADAR federated GW--radio follow-up framework~\cite{patel2025radarradioafterglowdetectionaidriven}. An extended validation of that original architecture is included in~\ref{app:attengw_validation}, while the present work and software release address the broader, architecture-independent detection pipeline.

\section{Methods}
\label{sec:methods}

We define the full detection pipeline and its components, then describe how they are selected and validated. We develop and tune our models using O3a development data, training on June–July 2019 noise and evaluating on August 2019 noise and a small set of catalog events. The resulting configuration is then trained on independent O3b data spanning December 2019 to January 2020, using January 28–February 29 as a held-out test set for final evaluation.

\subsection{Detection pipeline}

We present a framework for training and evaluating lightweight neural gravitational-wave detectors on real interferometer noise. The pipeline comprises five main components:\\
\hspace{6pt}
\begin{itemize}[label={}, leftmargin=0pt, itemsep=1pt, parsep=0pt, topsep=2pt]
\item \textbf{Downloader}: downloads strain data (signals and/or noise) from GWOSC and rejects or clamps glitchy segments.
    \item \textbf{Data generator}: constructs training segments from real noise and synthetic signals.
    \item \textbf{Model}: implements the selected neural architecture.
    \item \textbf{Training script}: handles optimization, validation, checkpointing, and run configuration.
    \item \textbf{Inference}: applies trained models to real detector data for inference.
\end{itemize}
\hspace{6pt}

We describe these components in more detail below. A shared configuration file is used for downloading and training, exposing most parameters intended for experimental variation while keeping preprocessing and training settings consistent across both stages. The configuration used for each training run is saved automatically with the corresponding outputs. A shorter, separate configuration is used for inference, containing only the settings needed for evaluation and trigger construction.

\subsubsection{Downloader}

To construct training and evaluation datasets, we developed a configurable downloader that retrieves Hanford and Livingston strain from the Gravitational Wave Open Science Center (GWOSC)~\cite{KAGRA:2023pio}. For user-selected GPS windows, we require times when both detectors are collecting. Training and test intervals are necessarily non-overlapping, with the training data preceding the test data in time. Training-noise data exclude padded regions around known events and undergo optional glitch interpolation followed by optional amplitude- and variance-based quality-control cuts. Test-noise data are retained without these cuts, with these same quality-control flags merely recorded for later analysis. Test-signal windows are centered on catalogued events and shortened when the full requested interval is unavailable.

A power spectral density is estimated using Welch's method~\cite{Welch1967TheUO}, from cleaned strain for training noise and raw strain for test data. Each window is stored as an HDF5 file containing the Hanford and Livingston strain, PSDs, frequency grid, and associated GPS, data-quality, and event metadata.


\subsubsection{Data generator}
\label{datagenerator}

To generate training and validation examples, we use a custom PyTorch dataloader that combines synthetic time-domain waveforms with real LIGO Hanford and Livingston noise. The noise samples are supplied by our downloader, whereas synthetic signals are supplied externally (e.g. through a surrogate waveform model or other waveform generator). For each sample, the dataloader either draws a pure-noise window or injects a waveform into a randomly selected stretch of strain from both detectors.

\begin{figure}[h!]
    \centering
    \includegraphics[width=\linewidth]{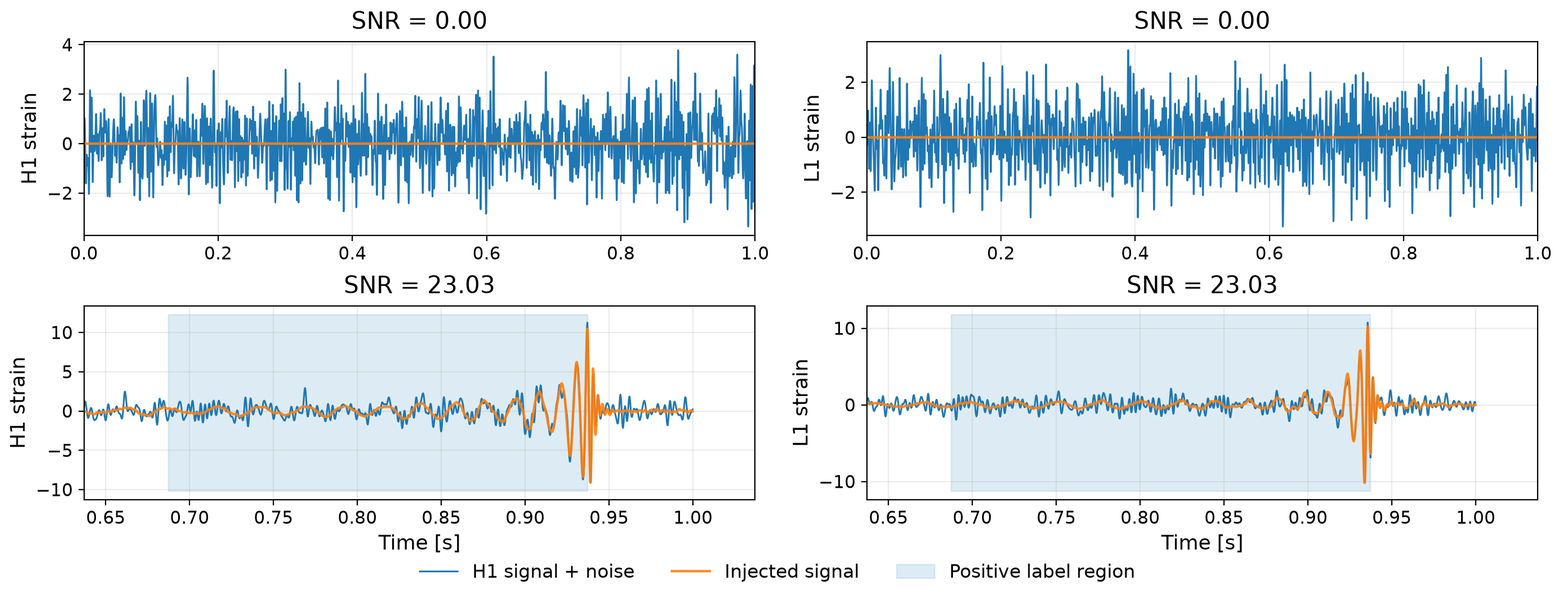}
        \caption{Example training samples. The plotted strain time series contain a binary black hole signal generated with \texttt{IMRPhenomXPHM}, injected into real LIGO noise from February 2020 (Hanford and Livingston). The data are whitened and cropped to a fixed-length window; the shaded region marks the timesteps labeled as positive (merger neighborhood).  Top row: Pure noise sample. Bottom row: scaled, high-SNR sample.}
    \label{fig:waveform-example}
\end{figure}

For injected samples, the signals are rescaled into a target SNR range and follow a simple curriculum~\cite{Louradour_curriculum}, in which higher SNRs are more likely in early epochs and gradually decrease over training. For the SNR evaluation, the PSDs are interpolated onto the FFT frequency grid and effectively band-limited to the LIGO sensitivity band (25–450 Hz), so that frequencies outside this range contribute negligibly. The signal is injected into raw strain, after which a configurable-length context is whitened using the corresponding PSDs and cropped to the fixed model window. The merger is placed randomly within the second half of this window. A binary target label is set to 1 in a configurable interval preceding the merger time and 0 elsewhere.  See Figure~\ref{fig:waveform-example} for an injected-signal and noise-only example from the data generator output. 

The input window length, label width, noise-only sampling probability, whitening context, SNR ranges and curriculum probabilities are configurable. The whitened Hanford and Livingston windows can additionally be normalized by a common scale derived from both detectors, preserving their relative amplitudes.

\subsubsection{Models}

To map fixed-length two-detector strain segments to per-timestep detection scores, we test a range of neural architectures. All models receive Hanford and Livingston strain with shape $(T,2)$ and return a scalar output $p^t\in[0,1]$ for each timestep. 

The models fall into four broad groups:

\begin{table}[]
\centering
\small
\setlength{\tabcolsep}{4pt}
\renewcommand{\arraystretch}{1.12}

\begin{tabularx}{\linewidth}{
    @{}
    >{\raggedright\arraybackslash}p{0.38\linewidth}
    >{\raggedright\arraybackslash}X
    @{}
}

\multicolumn{2}{@{}l}{\textbf{HDCN-based models}} \\[2pt]

HDCN + cross-attention (\textsc{AttenGW}) &
Per-detector HDCN with bidirectional cross-detector attention
\textit{(variants: without residual connection and/or LayerNorm)}. \\

HDCN + graph &
Per-detector HDCN with graph aggregation. \\

HDCN + temporal self-attention &
Separate HDCN detector encoders followed by learned fusion and temporal self-attention. \\[5pt]

\multicolumn{2}{@{}l}{\textbf{TCN fusion models}} \\[2pt]

TCN early fusion &
H and L combined before temporal feature extraction \textit{(variants: increased kernel size and hidden channels)}. \\

TCN separate stems + simple fusion &
Independent H/L TCN stems followed by concatenation and learned convolutional fusion. \\

TCN separate stems + interaction &
Independent H/L stems with product and absolute-difference interaction features. \\[5pt]

\multicolumn{2}{@{}l}{\textbf{TCN attention and representation models}} \\[2pt]

TCN + cross-attention &
Independent H/L TCN stems, downsampling, bidirectional cross-detector attention, and upsampling
\textit{(variant: full-resolution attention without downsampling)}. \\

TCN + temporal attention &
Early-fusion TCN followed by temporal self-attention
\textit{(variants: two attention heads; additional feed-forward layer)}. \\

TCN + gated temporal attention &
Early-fusion TCN followed by temporal self-attention with a learned gate controlling the attention contribution. \\

TCN + STFT &
Parallel time-domain TCN and STFT--Transformer branches
\textit{(variant: multi-resolution STFT branches)}. \\[5pt]

\multicolumn{2}{@{}l}{
\textbf{Other architectures}
} \\[2pt]

ResNet &
Dense one-dimensional ResNet with hierarchical downsampling and upsampling to recover per-timestep predictions. \\

BiMamba &
Local convolutional embedding followed by bidirectional residual Mamba blocks
\textit{(variant: unidirectional Mamba)}. \\

Patch Transformer &
Transformer operating on overlapping time-domain patches. \\

\end{tabularx}

\caption{Model architectures considered in this study. Parenthetical entries denote additional variants tested. H and L stand for Hanford and Livingston channels respectively}
\label{tab:models}
\end{table}

The HDCN models use separate WaveNet-style encoders~\cite{oord2016wavenet} for the Hanford and Livingston channels, following the HDCN setup in Ref.~\cite{Tian:2023vdc}. The graph model retains the graph-based aggregation from that architecture, while \textsc{AttenGW} replaces it with bidirectional cross-detector attention. The cross-attention operation uses scaled dot-product attention~\cite{vaswani2017attention} and the updated detector representations are followed by LayerNorm~\cite{ba2016layer}. We first introduced an early version of \textsc{AttenGW} as the gravitational-wave component of the RADAR federated GW–radio follow-up framework \cite{patel2025radarradioafterglowdetectionaidriven}. A variation on the attention module concatenates the two HDCN feature streams and applies temporal self-attention across the full sequence. A separate validation of the original \textsc{AttenGW} configuration against the graph-based February 2020 benchmark is presented in~\ref{app:attengw_validation}.

\begin{figure*}[p]
    \centering
    \includegraphics[width=\textwidth]
      {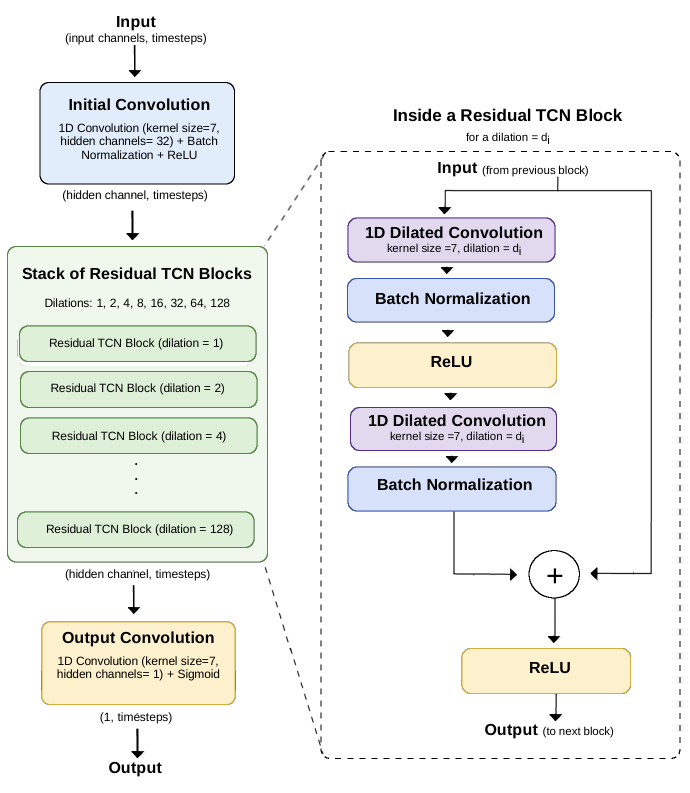}
    \caption{
    Schematic for the TCN early fusion model. In this study, we consider two input channels (one each for Hanford and Livingston) and 4096 timesteps.
    }
    \label{fig:schematic}
\end{figure*}

The TCN models use TCN-style residual blocks of dilated one-dimensional convolutions~\cite{bai2018empirical}. In the early-fusion model, Hanford and Livingston channels enter the first convolution together and are processed jointly throughout the TCN. For a schematic for the TCN early fusion model, see Figure~\ref{fig:schematic}. The separate-stem models instead use independent TCN encoders for Hanford and Livingston channels. The simple-fusion model concatenates the resulting representations before the output head, while the interaction model additionally includes their element-wise product and absolute difference, a construction also used in Ref.~\cite{conneau2017supervised}.

The remaining TCN models add cross-detector attention, temporal self-attention, gated temporal attention, or a parallel STFT representation to these convolutional encoders.

The ResNet-style model uses one-dimensional residual blocks~\cite{he2016deep} together with a feature-pyramid-style top-down pathway~\cite{lin2017feature} to combine features at multiple temporal resolutions. The BiMamba model applies Mamba selective state-space blocks~\cite{gu2024mamba} in forward and reversed temporal directions, following the bidirectional use of Mamba in Ref.~\cite{zhu2024visionmamba}. The Patch Transformer divides the two-detector time series into overlapping patches, following the patch-token representation used for time-series Transformers~\cite{nie2023timeseries}, and processes the resulting tokens with Transformer encoder layers~\cite{vaswani2017attention}.

During model development, we chose to not carry forward several variants because they converged more slowly and reached higher validation loss than the corresponding retained models: \textsc{AttenGW} without LayerNorm and/or residual attention connections, TCN early fusion with increased kernel size and hidden channels, full-resolution TCN cross-attention, TCN temporal attention with an additional feed-forward layer, multi-resolution STFT, unidirectional Mamba, and the Patch Transformer. Implementations of all models carried forward beyond this preliminary screening are included in the published repository.

\subsubsection{Training}
We train the selected model in PyTorch~\cite{paszke2019pytorchimperativestylehighperformance} and PyTorch Lightning~\cite{Falcon2019Lightning} using distributed data parallelism across available GPUs. The network is optimized with the Adam optimizer ~\cite{kingma2017adammethodstochasticoptimization}, using a binary cross-entropy loss between predicted scores and the binary labels defined in the data generator. A \texttt{ReduceLROnPlateau} scheduler monitors the validation loss and reduces the learning rate (lr) when progress stalls. The SNR seen by the models is rescaled into fixed target bands, with the injection mix gradually decaying from predominantly high-SNR samples to predominantly lower-SNR signals over a predetermined number of epochs. Training and validation use the same data-generation and whitening pipeline but draw simulated signals from separate user-specified files. The validation set uses a fixed SNR curriculum throughout training, and the checkpoint with the lowest validation loss is retained for evaluation.

\subsubsection{Inference}
\label{inference}

For inference, we load a trained model together with its automatically saved configuration and the Hanford and Livingston strain and PSDs produced by the downloader. The strain is divided into overlapping model-length windows using a configurable stride and set of offsets. Because mergers are placed in the latter half of windows during training, these inference settings should ensure that each candidate time appears in the latter half of at least one inference window (this is e.g. achieved for stride 4096 and offsets [0,2048]). Additional offsets can improve recovery by reducing window-alignment effects, but inference cost scales approximately linearly with their number. Each window is processed using the same whitening context, edge buffer, frequency range, and optional normalization used during training. 

The model produces a per-timestep score for each window, which requires further processing to be interpretable. To do so, we consider three trigger schemes. In \texttt{smoothed\_max}, the score is smoothed with a moving average and a trigger is recorded when its maximum exceeds the detection threshold, with the location of the maximum taken as the candidate time. In \texttt{peak\_width}, \texttt{scipy.signal.find\_peaks}~\cite{scipyfindpeaks} is applied to the raw score with a minimum peak height equal to the detection threshold and a configurable minimum peak width. \texttt{full\_peak} applies the same height and width requirements, followed by a body cut requiring more than half of the samples within the identified peak interval to exceed a secondary threshold. This criterion is loosely analogous to the additional mean-based cut used in Ref.~\cite{Tian:2023vdc}; however, whereas their secondary threshold is fixed, ours is defined dynamically relative to the primary threshold. Specifically, this secondary threshold is defined as the detection threshold minus a configurable margin, capped at a fixed maximum value. 

Candidate triggers from overlapping model windows are subsequently merged within a specified time tolerance. 

\paragraph{Noise similarity}
Changes in detector conditions can cause the distribution of evaluation noise to differ from that of the training data, representing a form of dataset or domain shift~\cite{2018arXiv181011953R,MORENOTORRES2012521}. Such shifts can degrade model performance and are therefore important to identify when evaluating robustness.

To characterize this shift in our setting, we introduce the
amplitude-spectral-density (ASD) mismatch diagnostic, which we compute as follows: Training and evaluation data are first divided into fixed-length segments. For each fixed-length segment \(s\), we define the detector-averaged log-ASD feature

$$
x_s(f_i)=\ln\!\left[\sqrt{A_{s,H}(f_i)A_{s,L}(f_i)}\right],
$$

where \(A_{s,H}\) and \(A_{s,L}\) are the Welch ASD estimates for Hanford and Livingston, and the ASD is defined as the square root of the corresponding power spectral density, \(A(f)=\sqrt{S(f)}\).

$$
d(s,t)=\frac{1}{N_f}\sum_{i=1}^{N_f}\left|x_s(f_i)-x_t(f_i)\right|.
$$

The ASD mismatch is then defined from the \(k\) nearest training segments as

$$
d_{\mathrm{ASD}}(s)=
\exp\!\left[
\frac{1}{k}\sum_{t\in\mathcal N_k(s)}d(s,t)
\right],
$$

where k is a user-defined parameter. Thus, \(d_{\mathrm{ASD}}=1\) indicates identical ASDs, while values above unity represent increasing spectral differences from the nearest training-noise segments. The evaluation mismatch distribution can then be divided into low-, medium-, and high-mismatch regimes, either by percentiles or absolute values.

This idea provides a general, model-agnostic way to attach a qualitative measure of domain shift—and thus some context for confidence in the resulting predictions—without interpreting the ASD mismatch as a calibrated uncertainty estimate. Estimating such shifts \textit{a priori} is important for assessing robustness, and the sensitive-volume curves in Sec.~\ref{sec:ablationstudies} show a general trend toward better performance for lower-ASD-mismatch segments, particularly at low false-positive counts, although the separation between mismatch regimes is not always large.

\subsection{Ablation experiments}

We evaluate a series of model, data-processing, training, signal, and inference ablations using sensitive volume as a function of false positives (FPs) on August 2019 O3a development data. We take the following configuration as a reference point for the ablation studies. Unless otherwise noted in the corresponding subsection, all ablations use this configuration, with only the parameter under study varied and all other parameters held fixed. The reference configuration uses a batch size of 32, an initial learning rate of $10^{-3}$, weight decay of $10^{-5}$, and a maximum of 75 epochs. The learning rate is reduced after two epochs without improvement in validation loss. Input windows contain 4096 samples ($1\,\mathrm{s}$), with a positive label width of 1024 samples ($250\,\mathrm{ms}$), and are whitened using 8\,s of surrounding noise. We enable per-window normalization. During training, noise-only examples are drawn with probability 0.6. For injected examples, the SNR curriculum transitions from a high-SNR range of $[10,25]$ to a low-SNR range of $[7,15]$, with the probability of drawing from the high-SNR range decreasing from 0.90 to 0.25 over training. We test a variety of trigger schemes, across which the following parameters are fixed: The score-smoothing width for \texttt{smoothed\_max} is 64 samples; inference uses 4096-sample windows (1\,s) with stride 4096 and offsets [0, 2048], giving roughly 50$\%$ overlap between the interleaved windows. Triggers within 0.25 seconds are merged, and an event counts as recovered if a trigger is within 0.25 seconds before or after the true merger time.

\begin{figure*}[h!]
\centering
\begin{subfigure}[t]{\linewidth}
\centering
\includegraphics[width=\linewidth]{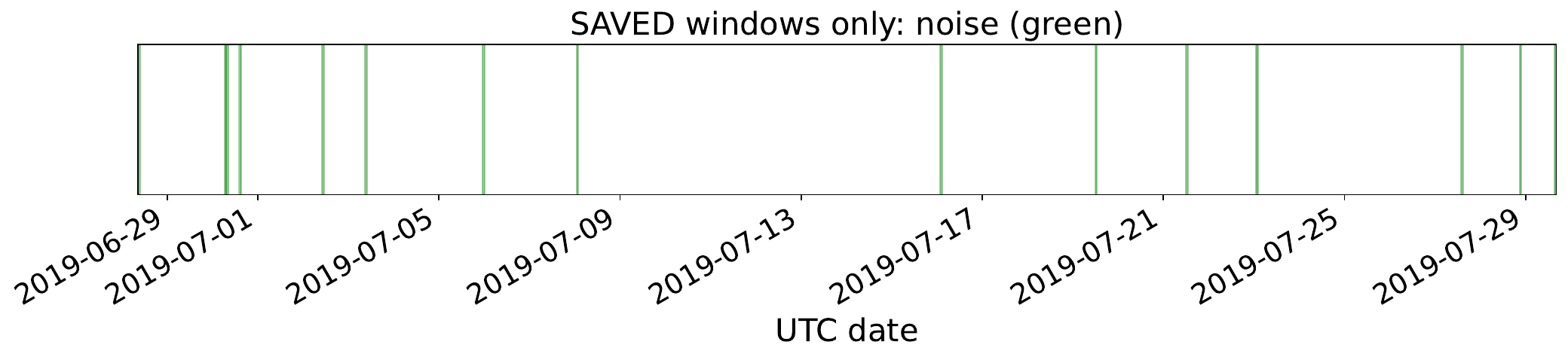}
\end{subfigure}

\begin{subfigure}[t]{\linewidth}
\centering
\includegraphics[width=\linewidth]{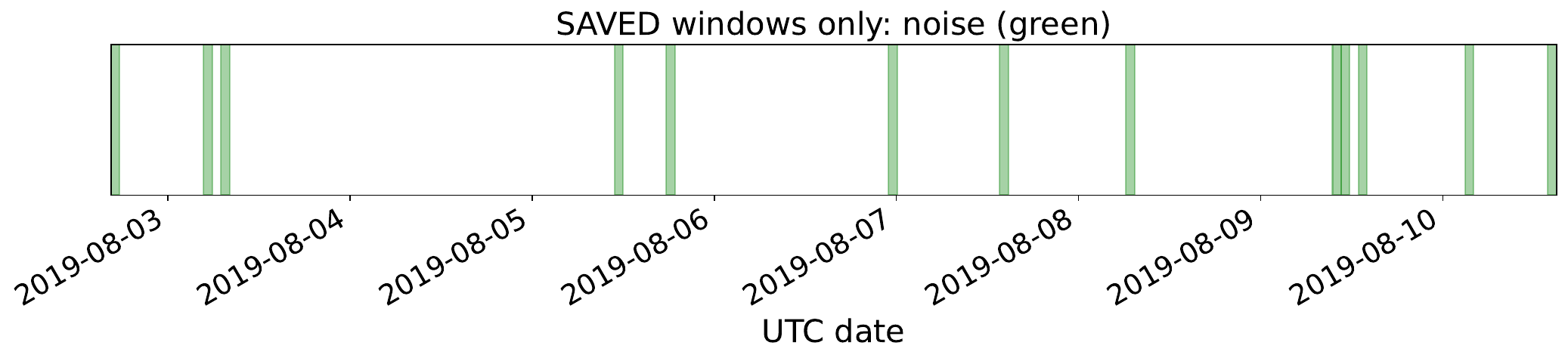}
\end{subfigure}
\caption{Saved Hanford--Livingston development training-noise segments (top) and development evaluation-noise segments (bottom) used for the ablation study}
\label{fig:segments}
\end{figure*}

The development training-noise dataset consists of 15 randomly selected 4096\,s Hanford--Livingston segments spanning June 28 to July 29, 2019 (UTC), for a total of 17.1\,h. The development evaluation-noise dataset contains 13 separate 4096\,s segments, selected randomly within a time span spanning August 1 to August 14, 2019 (UTC), totaling 14.8\,h. Figure~\ref{fig:segments} shows the individual saved segments. Noise is sampled at 4096\,Hz and analyzed between 25 and 450\,Hz. For the training-noise download we use a 30\,s event exclusion region, a glitch threshold of $5\sigma$, a maximum flagged-sample fraction of $10^{-3}$, a maximum Hanford/Livingston standard-deviation ratio of 10, and 4\,s Welch PSD segments. For background estimation, we use independently shifted Hanford and Livingston noise segments~\cite{W_s_2009} to increase the effective background livetime within the available GPS range. We verify that this time shifting does not materially change the false-positive behavior of the tested models (Appendix Fig.~\ref{fig:shifted_vs_unshifted}).

Synthetic BBH signals are generated with \texttt{IMRPhenomXHM}~\cite{PhysRevD.102.064002}. We use 500,000 waveforms for training, 10,000 waveforms for validation, and a separate set of 10,000 waveforms for the sensitive-volume evaluation. Component masses are drawn independently and uniformly from $[5,50],M_\odot$. Aligned spins are drawn uniformly from $[-0.8,0.8]$, and distances are sampled uniformly in volume between 100 and 1500\,Mpc. Evaluation waveforms are injected into the held-out August 2019 noise and detection efficiency $\epsilon(D)$ is measured in distance bins. We compute the corresponding sensitive volume,

\begin{equation}
V_{\mathrm{eff}}=\int \epsilon(D)\,4\pi D^2\,\mathrm{d}D,
\end{equation}
and pair it with the number of false-positive triggers obtained from the corresponding noise background at the same detection threshold. This follows the injection-based search evaluation used in MLGWSC-1, which reports the equivalent sensitive distance as a function of false-alarm rate ~\cite{Schaefer2023MLGWSC}. We report false-positive counts directly rather than converting them to a false-alarm rate.

For all ablations, the curves are evaluated separately for the three noise-similarity regimes defined in Sec.~\ref{inference}. To define these regimes, the training and evaluation noise are subdivided into 10\,min intervals and an ASD is computed for each interval. Each evaluation interval is then compared with its single nearest training interval ($k=1$), and the resulting mismatch distribution is divided at the 20th and 80th percentiles into low-, medium-, and high-mismatch regimes. The corresponding mismatch ranges are, respectively, \[ 1.00 \leq d_{\mathrm{ASD}}^{low} < 1.03,\qquad 1.03 \leq d_{\mathrm{ASD}}^{medium} < 1.20,\qquad 1.20 \leq d_{\mathrm{ASD}}^{high} \leq 1.32. \] After fixing these percentile-based boundaries, we sample the three regimes equally in all subsequent evaluations. Thus, a reported background livetime of one week for a plot means that each individual curve was generated by scoring the model on one week of background data drawn from the corresponding ASD-mismatch bin.

All models are trained for 75 epochs on a single NVIDIA A100 GPU, with the checkpoint attaining the lowest validation loss used for evaluation. Settings carried forward to subsequent ablations are selected primarily according to sensitive volume at the zero-false-positive operating point; for multi-seed experiments, we additionally consider the variation across seeds and for SNR experiments, we also consider performance on real signals. Where these criteria do not identify a clear preference, we compare the full sensitive-volume--versus--false-positive curves qualitatively.

\paragraph{Architecture and inference-trigger experiments}
We first compare the model architectures listed in Table~\ref{tab:models} (excluding the patch transformer because of its poor convergence performance). We use the reference training configuration, except that per-window normalization is disabled and the whitening-window length is set to 4s. We sweep the detection threshold and compare the \texttt{smoothed\_max}, \texttt{peak\_width}, and \texttt{full\_peak} trigger definitions. Width-based triggers are evaluated at 512 samples and 992 samples, corresponding respectively to half of the 1024-sample training label width and approximately the full label width with a small tolerance. For each model and inference configuration, each ASD-mismatch regime is evaluated using 5,000 signal injections and two weeks of shifted noise-background livetime. 

The direct benchmark comparison specifically between the original \textsc{AttenGW} model and the graph-based architecture is reported separately in \ref{app:attengw_validation}, as it uses a legacy preprocessing and evaluation configuration and is not used for model selection.

\paragraph{Multi-seed and normalization experiments}
For the two best-performing architectures from the initial screening, we compare training with and without shared per-window normalization. We use the reference training configuration. Whitening normalizes the frequency-dependent noise scale but does not explicitly fix the overall time-domain amplitude of each input. Shared per-window normalization may therefore stabilize training, while potentially removing absolute amplitude information that helps distinguish signals from noise. Five independent training seeds are evaluated for each architecture--normalization combination using the two trigger schemes retained from the preceding experiment. Each seed is evaluated using one week of noise-background livetime and 5,000 injections per ASD-mismatch regime. Four-seed repetitions for five additional architectures, without the normalization comparison, are reported in \ref{app:additional_seeds}.

\paragraph{Signal-family generalization}
For the best-performing architecture, we repeat the evaluation without retraining on precessing BBH and NSBH signals, using the best inference scheme retained from the architecture screening and the preferred normalization setting for each model. For each family, we use 5,000 new injections in each ASD-mismatch regime and otherwise retain the same evaluation procedure. The precessing-BBH set provides the primary signal-generalization test because it extends the aligned-spin BBH population used in training while retaining the same source class. The NSBH evaluation is treated as an exploratory out-of-distribution test because it also changes the characteristic masses, signal durations, tidal physics, and likely optimal training configuration.

Precessing BBH signals are generated with \texttt{IMRPhenomXPHM}~\cite{Pratten:2020ceb}. Component masses are drawn independently and uniformly from $[5,50],M_\odot$. Spin magnitudes are drawn uniformly from $[0,0.8]$ with isotropically distributed spin directions, and distances are sampled uniformly in volume between 100 and 1500\,Mpc.

NSBH signals are generated with \texttt{IMRPhenomNSBH}~\cite{thompson2020modelinggravitationalwavesignature}. Neutron-star masses are drawn uniformly from $[1.2,2.0],M_\odot$, with mass ratios $q=m_{\rm BH}/m_{\rm NS}$ drawn uniformly from $[4,7]$. The black-hole aligned spin is drawn uniformly from $[-0.5,0.75]$, while the neutron star is taken to be nonspinning. Its tidal deformability is drawn uniformly from $\Lambda_{\rm NS}\in[200,3000]$, and distances are sampled uniformly in volume between 20 and 500\,Mpc.

\paragraph{Training and preprocessing experiments}
We additionally vary the learning-rate scheduler, training-label width, and whitening-context length. The reference scheduler uses \texttt{ReduceLROnPlateau} with a reduction factor of 0.5, patience of two epochs, an absolute improvement threshold of $3\times10^{-4}$, and a minimum learning rate of $10^{-6}$. We compare this with a less responsive scheduler using a patience of three epochs and an absolute threshold of $10^{-5}$, with the remaining settings unchanged. We compare the reference label width of 1024 samples (250\,ms) with a shorter width of 512 samples (125\,ms). An additional soft cosine label, which replaces the abrupt transition from 0 to 1 with a cosine-shaped ramp preceding the positive interval, was tested during development but excluded from the quantitative comparison because it did not converge reliably. Finally, we compare whitening contexts of 4, 8, 16, and 32\,s, with 8\,s used as the reference setting. Each experiment changes only one of these settings, allowing the same reference run to serve as the baseline for all three comparisons. Each comparison uses a single training seed and two weeks of noise-background livetime per curve.

We also compare the default glitch-handling configuration with stricter downloader-level rejection and per-window rejection during training, testing whether reducing the diversity of transient noise morphologies simplifies the learning problem and improves convergence.

\paragraph{SNR curriculum and real signal inference}

We evaluate the sensitivity of the pipeline to the SNR distribution used during training through a broad curriculum ablation. The tested configurations vary the low- and high-SNR sampling ranges, the fraction of noise-only examples, and the schedule controlling the probability of drawing from the higher-SNR range over the course of training. The full set of varied SNR parameters is summarized in Table~\ref{tab:snr_ablation_real_smoothed_max}. Each configuration is trained with a single seed and first evaluated through an injection study using two weeks of shifted noise-background livetime per sensitive-volume curve.

To test whether the injection-study trends carry over to observed signals, we additionally evaluate eight curriculum configurations on eight cataloged gravitational-wave events from August 2019 as well as the same 13 August 2019 noise files employed in the preceding ablation study, corresponding to approximately 20 hours of real detector data in total.

\subsection{Benchmark on real O3b data}
\label{methods:realdata}

For the final O3b evaluation on real noise and signals, we use the model, normalization, training, and inference settings selected from the preceding ablation experiments on O3a development data. The supporting comparisons are presented in Sec.~\ref{sec:ablationstudies}.

The training-noise set consists of 100 randomly selected 4096\,s coincident Hanford--Livingston segments from December 1, 2019, through January 15, 2020, followed by all available coincident segments from January 15 through January 27, 2020, for a total coincident livetime of 12.2 days. The final evaluation set spans January 28 through February 29, 2020, ending on March 1 at 00:00 UTC, and contains approximately 22.7 days of coincident livetime. In addition to the coincident noise data, we download the 13 confident GWTC-3 events~\cite{KAGRA:2021vkt} occurring during this interval. These data are used for the final injection-based sensitivity study and for evaluating false positives in real detector noise and recovery of cataloged signals.

For training and sensitivity evaluation injections, synthetic BBH signals are generated with \texttt{IMRPhenomXHM}. We use 800,000 waveforms for training, 50,000 for validation, and 10,000 for injection-based sensitivity evaluation. Detector-frame component masses are sampled from $m_1\in[5,95],M_\odot$ and $m_2\in[2.5,55],M_\odot$, with $q\leq10$ and total mass below $140\,M_\odot$. Aligned spins are drawn uniformly from $[-0.8,0.8]$, and distances are sampled uniformly in volume between 100 and 1500\,Mpc.

We train four early-fusion TCNs with varying seeds, using a batch size of 32, an initial learning rate of $10^{-3}$, and a maximum of 75 epochs. Input segments contain 4096 samples, with a positive label width of 1024 samples and an edge buffer of 2048 samples. Each input is whitened using an 8\,s context and then normalized using a shared per-window scale across both detectors. Noise-only examples are drawn with probability 0.65. We use low- and high-SNR ranges of $[3,10]$ and $[10,25]$, with $p_{\mathrm{high}}$ decreasing from 0.90 to 0.25. Based on the O3a development study, we also fix the \texttt{smoothed\_max} threshold at 0.9999. Consistent with the ablation studies, we select the checkpoint with the lowest validation loss for evaluation. We use a score-smoothing width of 64 samples for \texttt{smoothed\_max}; triggers within 0.25 seconds are merged, and an event counts as recovered if a trigger is within 0.25 seconds before or after the true merger time. For the injection-based sensitivity study, inference uses 4096-sample windows with stride 4096 and offsets \([0,2048]\). To reduce sensitivity to window alignment, recovery of cataloged signals and evaluation of the 22.7-day real-noise background instead use offsets \([0,1024,2048,3072]\); the smaller offset set keeps the substantially larger injection study computationally tractable.

 The early-fusion TCN is a lightweight model with only 116,000 trainable parameters. Training is performed on one NVIDIA H100 GPU on the DeltaAI cluster and completes in under nine hours for 75 epochs -- though convergence is typically reached within the first 30 epochs. Inference and trigger construction on the held-out O3b evaluation set, comprising 22.7 days of noise livetime and 13 confident GWTC-3 events, was completed in under 5.0 hours on one NVIDIA A100 GPU on the Delta cluster -- about 100x faster throughput than real time\footnote{Cataloged-signal recovery and real-noise evaluation use four offsets rather than the two used for the injection study, approximately doubling the inference time.}. For the synthetic injection study, we construct and evaluate 365 days of effective noise livetime, which we complete in under 41 hours on one NVIDIA A100 GPU on the Delta cluster.

For the O3b analyses, let \(N\) denote the number of false-positive triggers
observed over a background livetime \(T\). We model \(N\) as a stationary
Poisson process, appropriate for independent events occurring at an
approximately constant rate~\cite{poissongarwood},
\[
P(N=n\mid\mu)=\frac{\mu^n e^{-\mu}}{n!},
\]
where \(\mu\) is the expected number of triggers during \(T\). For zero observed triggers, the one-sided 95\% upper limit
\(\mu_{95}\) is defined by
\[
P(N=0\mid\mu_{95})=e^{-\mu_{95}}=0.05,
\]
giving \(\mu_{95}=-\ln(0.05)\simeq2.996\). The corresponding upper limit on the
false-alarm rate is therefore \(r_{95}=\mu_{95}/T\).

\section{Results}
\subsection{Ablation studies}
\label{sec:ablationstudies}

\subsubsection{Architecture and inference-trigger comparison}

We first compared the three trigger definitions and the two widths used for the width-based methods. Table~\ref{tab:inference_by_asd} reports performance when the threshold is selected independently in each ASD-mismatch bin as the lowest tested value with zero observed false positives. The \texttt{smoothed\_max} trigger provided such an operating point for 11 of the 12 models in every bin, more than any other, and yielded the highest median sensitive volume in the low- and medium-mismatch bins, although individual width-based configurations attained larger maximum values for particular models. The \texttt{peak\_width} and \texttt{full\_peak} results are identical at a fixed width, indicating that the additional mean-score requirement in \texttt{full\_peak} does not meaningfully alter recovery. The two widths perform similarly in the low- and medium-mismatch bins, suggesting that scores elevated over half the training-label width are often elevated over nearly its full width. The 992-sample width is nevertheless more robust in the high-mismatch comparisons.

\begin{table*}[h!]
\centering
\normalsize
\renewcommand{\arraystretch}{1.1}
\setlength{\tabcolsep}{2pt}

\begin{tabularx}{\textwidth}{llcccc>{\raggedright\arraybackslash}X}
\toprule
Inference trigger &  \shortstack{$d_{ASD}$} & \shortstack{0 FP\\models} & \shortstack{Median\\threshold} & \shortstack{Median\\$V_{\mathrm{eff}}$} & \shortstack{Maximum\\$V_{\mathrm{eff}}$} & \shortstack{Best\\model}\\
\midrule
smoothed\_max & Low & 11/12 & 0.99999 & 0.914 & 2.779 & EF \\
smoothed\_max & Medium & 11/12 & 0.99999 & 0.897 & 2.695 & STFT \\
smoothed\_max & High & 11/12 & 0.99999 & 0.360 & 2.602 & STFT \\
peak\_width\_512 & Low & 10/12 & 0.999999 & 0.534 & 3.054 & STFT \\
peak\_width\_512 & Medium & 9/12 & 0.999999 & 0.450 & 2.635 & STFT \\
peak\_width\_512 & High & 10/12 & 0.9999945 & 0.281 & 1.769 & EF \\
peak\_width\_992 & Low & 10/12 & 0.999999 & 0.534 & 3.048 & STFT \\
peak\_width\_992 & Medium & 10/12 & 0.999999 & 0.473 & 2.629 & STFT \\
peak\_width\_992 & High & 10/12 & 0.99999 & 0.580 & 2.476 & SSF \\
full\_peak\_512 & Low & 10/12 & 0.999999 & 0.534 & 3.054 & STFT \\
full\_peak\_512 & Medium & 9/12 & 0.999999 & 0.450 & 2.635 & STFT \\
full\_peak\_512 & High & 10/12 & 0.9999945 & 0.281 & 1.769 & EF \\
full\_peak\_992 & Low & 10/12 & 0.999999 & 0.534 & 3.048 & STFT \\
full\_peak\_992 & Medium & 10/12 & 0.999999 & 0.473 & 2.629 & STFT \\
full\_peak\_992 & High & 10/12 & 0.99999 & 0.580 & 2.476 & SSF \\
\bottomrule
\end{tabularx}
\caption{Aggregate inference-trigger performance evaluated separately within each ASD-mismatch bin. For each model, trigger, and ASD bin, the sensitive volume, in Gpc$^3$, is evaluated at the lowest tested threshold with zero observed false positives in that bin. The best model is that corresponding to maximum $V_{\mathrm{eff}}$. SSF, EF, and STFT denote \texttt{TCN separate stems + simple fusion}, \texttt{TCN early fusion}, and \texttt{TCN + STFT}, respectively.}
\label{tab:inference_by_asd}
\end{table*}

\begin{figure*}[p]
    \centering
    \includegraphics[width=\textwidth]
      {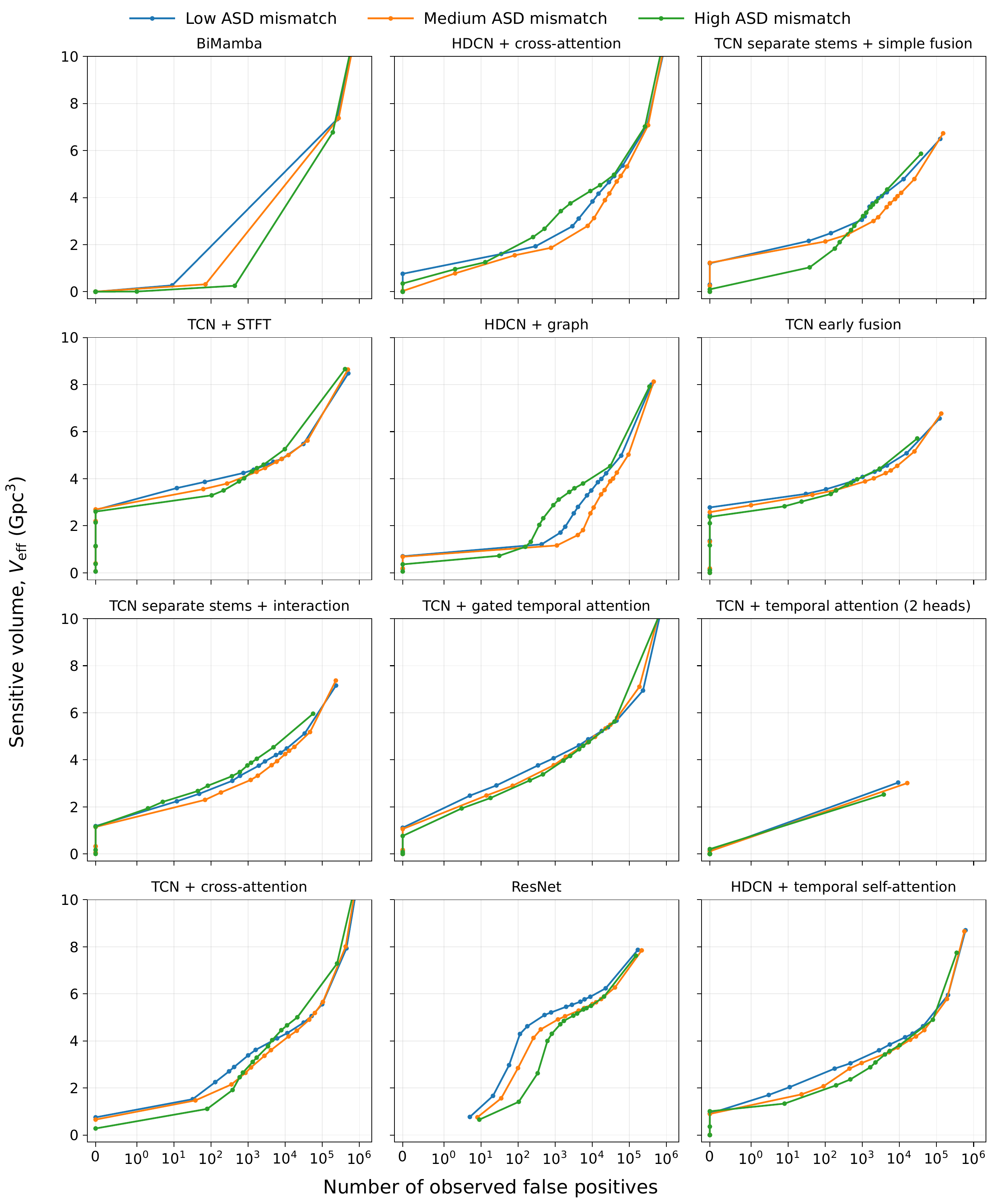}
        \caption{
        Sensitive volume as a function of the number of observed false positives for all 12 architectures using the \texttt{smoothed\_max} trigger. Each panel corresponds to one architecture, with separate curves for the low-, medium-, and high-ASD-mismatch bins. Each curve uses 5,000 aligned-spin BBH injections per bin, with component masses in $[5,50]\,M_\odot$, aligned spins in $[-0.8,0.8]$, and trained with an SNR curriculum transitioning from \([10,25]\) to \([7,15]\), together with two weeks of shifted noise background. The generally lower sensitive volume in the higher-mismatch bins demonstrates an empirical association with reduced sensitivity.
        }
    \label{fig:architecture_smoothed_max}
\end{figure*}

Notably, for \texttt{smoothed\_max}, the median sensitive volume decreased from $0.914$ to $0.897$ and $0.360,\mathrm{Gpc}^3$ across the low-, medium-, and high-mismatch bins, respectively. This ordering is also visible for the other trigger definitions and in the model-level curves, indicating that the ASD-mismatch bins capture increasing reconstruction difficulty, as intended.

Figure~\ref{fig:architecture_smoothed_max} shows the full sensitive-volume--versus--false-positive curves for all 12 architectures under \texttt{smoothed\_max}. Based on the aggregate results and the full model-level curves, we retain \texttt{smoothed\_max} as the primary trigger, though we also report results for \texttt{peak\_width} with a 992-sample width for select ablations moving forward. For the models, we retain \texttt{TCN + STFT} and \texttt{TCN early fusion} for the subsequent comparisons.

\begin{figure*}[!h]
    \centering
    \includegraphics[width=\textwidth]
    {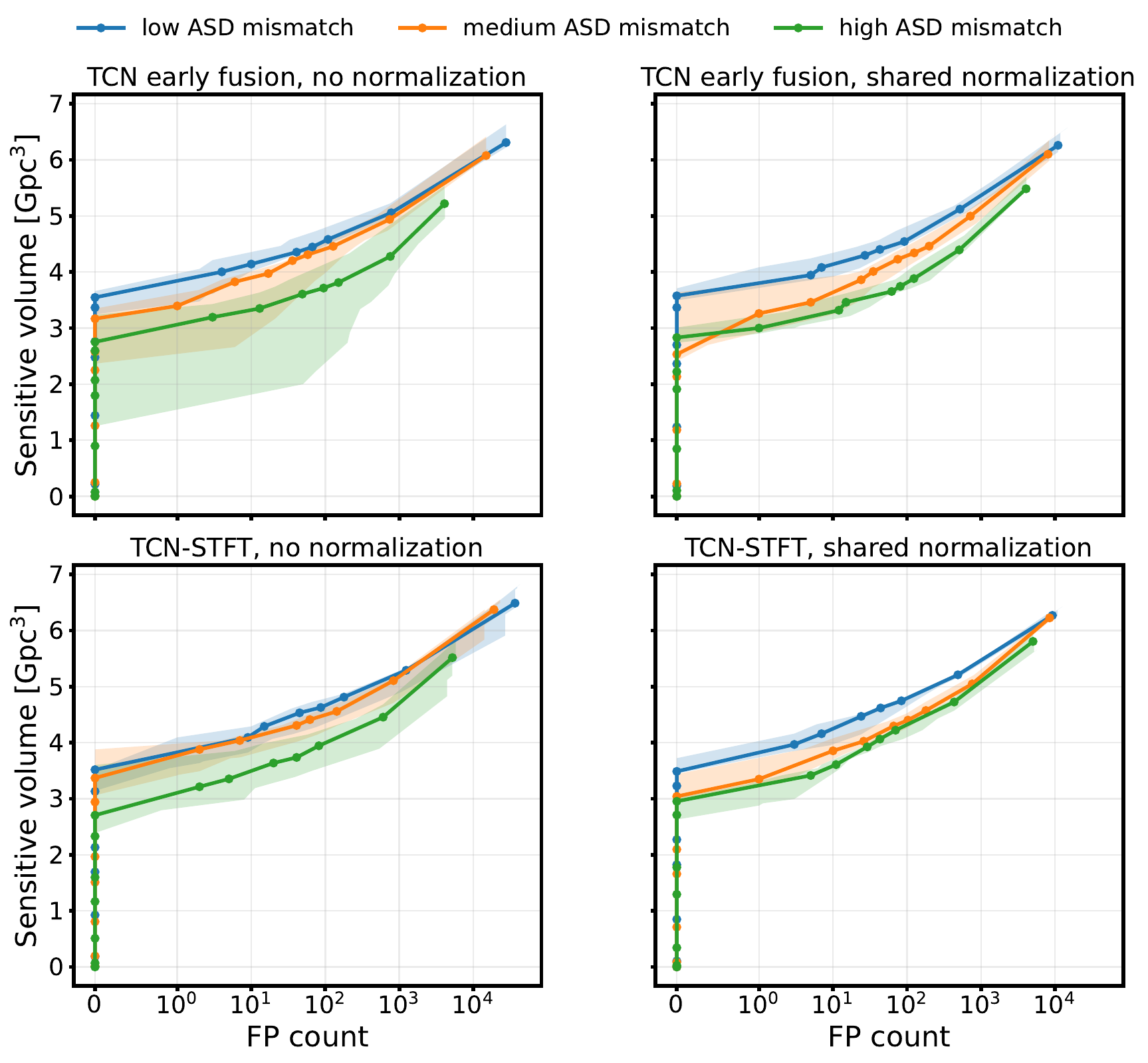}
    \caption{
    Sensitive volume as a function of the number of observed false positives under the \texttt{smoothed\_max} trigger across training seeds for the two retained architectures, with and without shared per-window normalization. The injection population is the same as in Figure~\ref{fig:architecture_smoothed_max}, with 5,000 injections and one week of shifted noise background per ASD-mismatch bin. Each panel shows the median across seeds for each threshold, with shaded bands indicating the full seed-to-seed range.
    }
    \label{fig:multiseed_normalization}
\end{figure*}

These comparisons are intended to evaluate practical lightweight implementations within a common detection pipeline, rather than to establish a general ranking of the underlying architecture families. Each family admits substantial variation in hyperparameters and the implementations tested here were not exhaustively optimized on an architecture-specific basis. The results should therefore be interpreted as applying primarily to the training procedure, data, and preprocessing choices. The comparatively weak performance of the tested ResNet may partly reflect differences in task formulation: our model produces per-timestep predictions, whereas many established models such as Aframe use a modified ResNet to map each fixed-length strain window to a scalar detection statistic~\cite{Marx2025Aframe}.

\subsubsection{Multi-seed and normalization experiments}

Figure~\ref{fig:multiseed_normalization} compares the two leading architectures across five independent training seeds, both with and without shared per-window normalization, using the \texttt{smoothed\_max} trigger. The shaded regions trace the pointwise minimum-to-maximum envelope of the five seed curves. To construct the median curve, we consider each tested threshold and take the median false-positive count and median sensitive volume across the five seeds. The resulting spread shows substantial seed-to-seed variation, particularly for TCN early fusion without normalization. Normalization generally narrows this spread, especially for high ASD mismatch $d_{ASD}$. At zero false-positive counts, TCN early fusion is marginally better in the low-ASD-mismatch regime, while the two architectures are otherwise closely matched, with TCN-STFT showing at most a slight advantage in the medium- and high-mismatch regimes. We retain normalization for its greater training stability and reduced seed sensitivity.

Four-seed repetitions for five additional architectures are reported in \ref{app:additional_seeds}. The additional multi-seed experiments show that the HDCN-based architectures remain substantially weaker than the TCN models. In contrast, the TCN separate-stem interaction and gated temporal-attention models are both competitive with TCN early fusion, with only modest differences across ASD-mismatch bins. This suggests that performance varies little among the strongest TCN architectures. We therefore retain TCN early fusion as the baseline because it achieves comparable performance with the simplest architecture.

\subsubsection{Signal-family generalization.}

\begin{figure*}[!h]
    \centering
    \includegraphics[width=\textwidth]
    {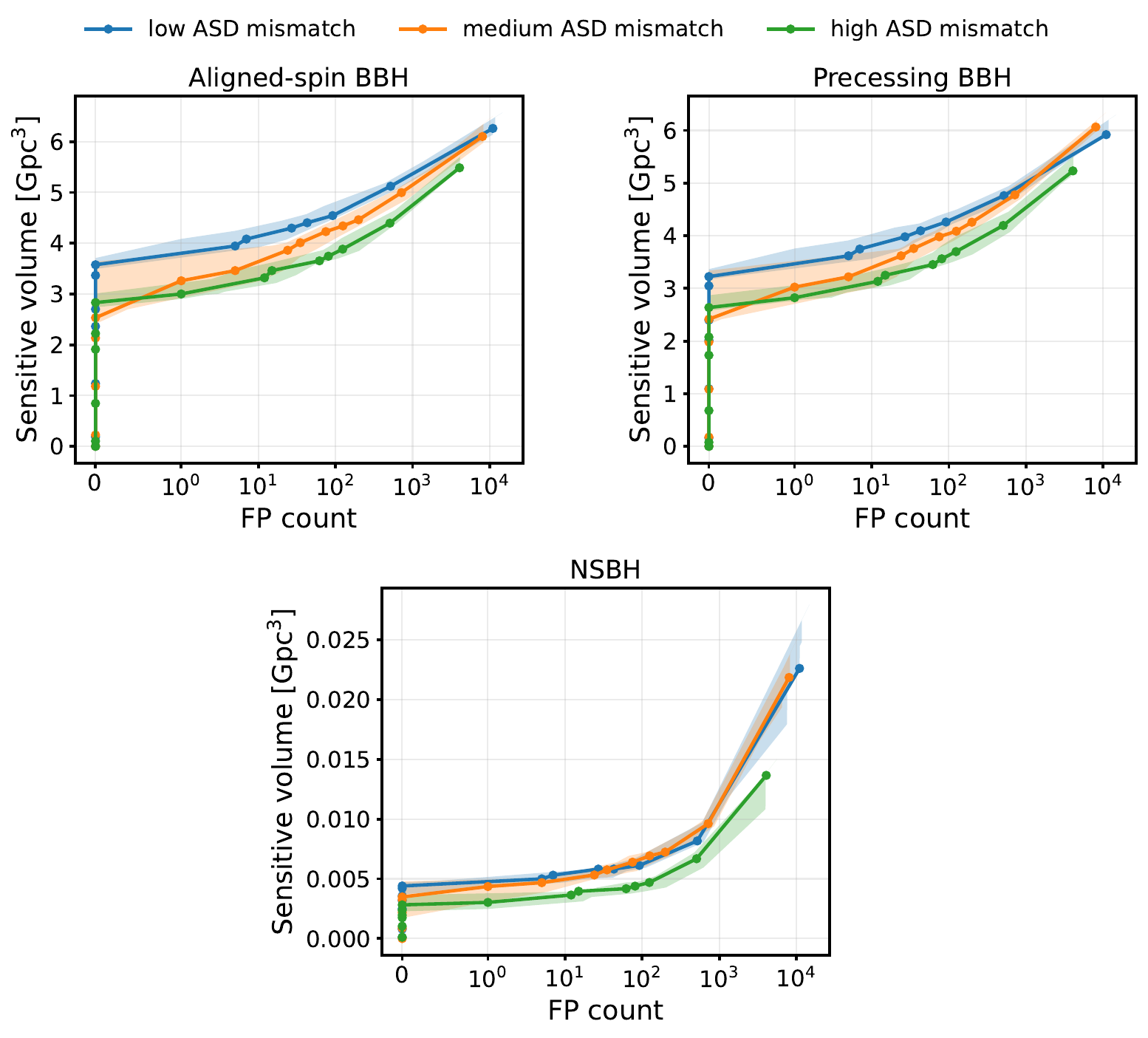}
    \caption{
    Sensitive volume as a function of false-positive count under the
    \texttt{smoothed\_max} trigger for the same trained models evaluated on
    aligned-spin BBH, precessing BBH, and NSBH populations. All models were trained with an SNR curriculum transitioning from \([10,25]\) to \([7,15]\). The BBH populations
    have component masses in $[5,50]\,M_\odot$; aligned spins are drawn from
    $[-0.8,0.8]$, while precessing spin magnitudes are drawn from $[0,0.8]$
    with isotropic orientations. The NSBH population has neutron-star masses
    in $[1.2,2.0]\,M_\odot$, mass ratios $q\in[4,7]$, black-hole aligned spins
    in $[-0.5,0.75]$, and nonspinning neutron stars. Each curve uses 5,000
    injections and one week of shifted noise background per ASD-mismatch bin.
    }
    \label{fig:signal_generalization}
\end{figure*}

Figure~\ref{fig:signal_generalization} compares the retained TCN early-fusion model on the aligned-spin BBH reference population, precessing BBH signals, and NSBH signals. The BBH reference population was generated from the same parameters as the training set. Recovery on the precessing-BBH population remains close to that of the aligned-spin reference population, with a modest reduction that is most apparent in the low-ASD-mismatch regime. Thus, the same aligned-spin-trained model may be performant on precessing BBHs without retraining or changing the inference procedure, so there is no additional computational cost. In contrast, the NSBH curves differ substantially, reflecting both the distinct waveform population and its different injection-volume range; their absolute sensitive volumes should therefore be interpreted within that population rather than compared directly with the BBH values.

\subsubsection{Training and preprocessing experiments.}

Table~\ref{tab:lr_dim_comparison} shows that the reference configuration outperforms both alternatives across all ASD-mismatch bins and trigger schemes. The less responsive learning-rate scheduler produces a moderate reduction in sensitive volume, while the shorter 512-sample label (as opposed to the 1024-sample reference label) is particularly detrimental for the \texttt{peak\_width} trigger.

\begin{table}
    \centering
    \begin{tabular}{llrrrr}
        \toprule
        Configuration & Trigger & Threshold & \multicolumn{3}{c}{$V_{\mathrm{eff}}$ [Gpc$^3$]} \\
        \cmidrule(lr){4-6}
        & & & $\boldsymbol{d_{\mathrm{ASD}}^{\mathrm{low}}}$
&
$\boldsymbol{d_{\mathrm{ASD}}^{\mathrm{medium}}}$
&
$\boldsymbol{d_{\mathrm{ASD}}^{\mathrm{high}}}$ \\
        \midrule
        Reference & \texttt{smoothed\_max} & 0.995 & 3.513 & 3.594 & 2.843 \\
        Reference  & \texttt{peak\_width} & 0.9995 & 3.368 & 3.507 & 2.668 \\
        Alt. LR schedule & \texttt{smoothed\_max} & 0.9999 & 2.741 & 2.722 & 2.170 \\
        Alt. LR schedule & \texttt{peak\_width} & 0.99999 & 2.282 & 2.251 & 1.748 \\
        512-sample label & \texttt{smoothed\_max} & 0.9995 & 2.883 & 2.939 & 2.412 \\
        512-sample label & \texttt{peak\_width} & 0.999999 & 0.590 & 0.615 & 0.340 \\
        \bottomrule
    \end{tabular}
    \caption{Alternative learning-rate-scheduler and training-label-width comparison. Sensitive volumes are evaluated at each model's lowest tested threshold with zero observed false positives in the low- and medium-ASD-mismatch backgrounds. The \texttt{peak\_width} trigger width is the training-label width minus 32 samples. The columns $d_{\mathrm{ASD}}^{\mathrm{low}}$, $d_{\mathrm{ASD}}^{\mathrm{medium}}$, and $d_{\mathrm{ASD}}^{\mathrm{high}}$ denote the low-, medium-, and high-ASD-mismatch regimes, respectively.}
    \label{tab:lr_dim_comparison}
\end{table}

\begin{table}[!h]
    \centering
    \begin{tabular}{llrrrr}
        \toprule
        Whitening context & Trigger & Threshold & \multicolumn{3}{c}{$V_{\mathrm{eff}}$ [Gpc$^3$]} \\
        \cmidrule(lr){4-6}
        & & & $\boldsymbol{d_{\mathrm{ASD}}^{\mathrm{low}}}$
&
$\boldsymbol{d_{\mathrm{ASD}}^{\mathrm{medium}}}$
&
$\boldsymbol{d_{\mathrm{ASD}}^{\mathrm{high}}}$ \\
        \midrule
        4 s & \texttt{smoothed\_max} & 0.99995 & 2.507 & 2.432 & 1.802 \\
        4 s & \texttt{peak\_width} & 0.999999 & 0.491 & 0.482 & 0.186 \\
        8 s (reference) & \texttt{smoothed\_max} & 0.995 & 3.462 & 3.548 & 2.797 \\
        8 s (reference) & \texttt{peak\_width} & 0.9995 & 3.351 & 3.414 & 2.659 \\
        16 s & \texttt{smoothed\_max} & 0.9999 & 2.856 & 2.958 & 2.412 \\
        16 s & \texttt{peak\_width} & 0.99999 & 2.164 & 2.213 & 1.690 \\
        32 s & \texttt{smoothed\_max} & 0.99995 & 2.740 & 2.883 & 2.241 \\
        32 s & \texttt{peak\_width} & 0.99999 & 2.618 & 2.753 & 2.048 \\
        \bottomrule
    \end{tabular}
    \caption{Whitening-context-length comparison. Sensitive volumes are evaluated at each model's lowest tested threshold with zero observed false positives in the low- and medium-ASD-mismatch backgrounds. The \texttt{peak\_width} trigger width is the training-label width minus 32 samples.}
    \label{tab:whitening_context_comparison}

\end{table}

Table~\ref{tab:whitening_context_comparison} shows that the 8\,s reference context achieves the highest sensitive volume in every ASD-mismatch bin under both trigger definitions. Both shorter and longer contexts reduce performance, indicating an intermediate optimum rather than a monotonic benefit from increasing the whitening duration. Longer whitening contexts also increase training and inference cost, making the strong performance of the 8\,s setting favorable from both accuracy and computational-efficiency perspectives.

In addition to these more stringent tests, we also compare the default glitch-handling configuration with stricter downloader-level rejection and per-window rejection during training. Since glitch rejection schemes are meant to stabilize training, we evaluate performance improvement solely through validation loss curves. Both stricter downloader-level rejection and per-window rejection led to slower convergence. We therefore retain a broader range of glitch morphologies in the training noise, allowing the models to learn directly from these nonstationary features.

\subsubsection{SNR curriculum and real signal inference}

We next examine how the training SNR curriculum affects both injection-based sensitivity and recovery of cataloged signals in real detector data.

\begin{figure*}[!h]
    \centering
    \includegraphics[width=\textwidth]
    {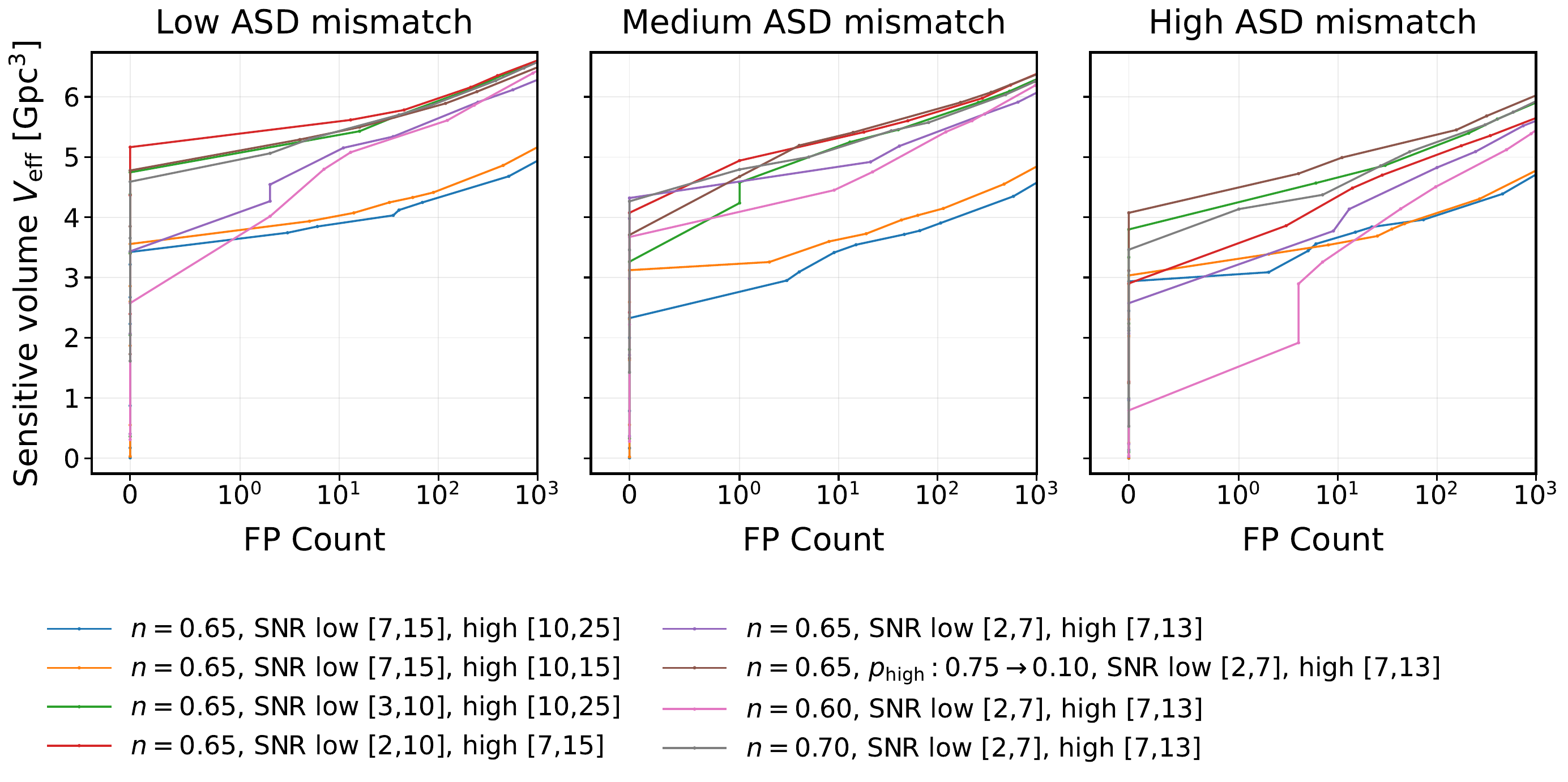}
    \caption{Sensitive volume as a function of false-positive count under the \texttt{smoothed\_max}  trigger for models trained with various SNR curricula.
    }
    \label{fig:snr_injections}
\end{figure*}

The injection study in Figure~\ref{fig:snr_injections} shows a clear dependence on the SNR distribution used during training. The two highest-SNR curricula, with ranges $([7,15]/[10,25])$ and $([7,15]/[10,15])$, consistently produce among the lowest sensitive-volume curves across all three ASD-mismatch bins. 

However, concentrating training within the narrower $[2,7]/[7,13]$ ranges is not uniformly effective. Its performance depends strongly on the noise-only fraction and on the decay of $p_{\mathrm{high}}$: it performs poorly for $p_{\mathrm{noise}}=0.60$, improves for $p_{\mathrm{noise}}=0.65$, and becomes competitive only when paired with either a more aggressive $p_{\mathrm{high}}$ decay or a larger noise-only fraction. The broader $[2,10]/[7,15]$ curriculum achieves the highest sensitivity in the low-mismatch regime but degrades more substantially as the ASD mismatch increases. 

The $[3,10]/[10,25]$ curriculum provides the strongest overall balance. It remains among the leading configurations across all three mismatch regimes while combining exposure to low-SNR signals with a broad higher-SNR component. The injection study therefore favors curricula that extend below the eventual detection limit without concentrating training too narrowly at the lowest SNRs.


To determine whether these trends generalize to observed signals, we evaluate the same curriculum configurations on eight cataloged gravitational-wave events and on the August 2019 development-noise background. For each configuration, Table~\ref{tab:snr_ablation_real_smoothed_max} reports the lowest tested \texttt{smoothed\_max} threshold yielding zero observed false positives and the corresponding number of recovered events.

\begin{table}[!h]
    \centering

    \begin{tabular}{cccccc}
        \toprule
        \multicolumn{4}{c}{Training configuration}
        & \multicolumn{2}{c}{Best 0-FP operating point} \\
        \cmidrule(lr){1-4}
        \cmidrule(lr){5-6}
        $p_{\mathrm{noise}}$
        & \shortstack{Low-SNR\\range}
        & \shortstack{High-SNR\\range}
        & $p_{\mathrm{high}}^{\mathrm{init}}
           \rightarrow p_{\mathrm{high}}^{\mathrm{final}}$
        & Threshold
        & Recovered \\
        \midrule

        $0.65$
        & $[3,10]$
        & $[10,25]$
        & $0.90 \rightarrow 0.25$
        & $0.9999$
        & $3/8$ \\

        $0.65$
        & $[2,7]$
        & $[7,13]$
        & $0.90 \rightarrow 0.25$
        & $0.99995$
        & $3/8$ \\

        \addlinespace[0.6em]

        $0.65$
        & $[2,10]$
        & $[7,15]$
        & $0.90 \rightarrow 0.25$
        & $0.99999$
        & $2/8$ \\

        $0.65$
        & $[2,7]$
        & $[7,13]$
        & $0.75 \rightarrow 0.10$
        & $0.999999$
        & $2/8$ \\

        $0.60$
        & $[2,7]$
        & $[7,13]$
        & $0.90 \rightarrow 0.25$
        & $0.99999$
        & $2/8$ \\

        $0.70$
        & $[2,7]$
        & $[7,13]$
        & $0.90 \rightarrow 0.25$
        & $0.99999$
        & $2/8$ \\

        \addlinespace[0.6em]

        $0.65$
        & $[7,15]$
        & $[10,25]$
        & $0.90 \rightarrow 0.25$
        & $0.99995$
        & $1/8$ \\

        $0.65$
        & $[7,15]$
        & $[10,15]$
        & $0.90 \rightarrow 0.25$
        & $0.99995$
        & $1/8$ \\

        \bottomrule
    \end{tabular}
    \caption{
        Real-data performance of the SNR-curriculum configurations using
    the \texttt{smoothed\_max} trigger. For each configuration, the
    reported operating point is the lowest tested threshold yielding
    zero observed false positives in the August 2019 development
    background.
    }
    \label{tab:snr_ablation_real_smoothed_max}
\end{table}

Table~\ref{tab:snr_ablation_real_smoothed_max} reinforces the injection-study result that training must include signals well below the eventual detection threshold. The two configurations recovering three events both expose the network to low SNRs in the low-SNR range, whereas the curricula restricted to $[7,15]$ recover only one of eight events. The $[3,10]/[10,25]$ and $[2,7]/[7,13]$ configurations each recover GW190814\_211039, GW190828\_063405, and GW190828\_065509, whose cataloged network matched-filter SNRs are $25.3^{+0.1}_{-0.2}$, $16.5^{+0.2}_{-0.3}$, and $10.2^{+0.4}_{-0.5}$, respectively~\cite{Abbott2024GWTC21}.

We therefore retain the $p_{\mathrm{noise}}=0.65$, $[3,10]/[10,25]$ curriculum with $p_{\mathrm{high}}:0.90\rightarrow0.25$. It is consistently competitive across the injection-study mismatch regimes and ties for the strongest real-signal recovery. Its corresponding zero-false-positive \texttt{smoothed\_max} operating point, 0.9999, is fixed for the final O3b evaluation. We also observe sensitivity to inference-window alignment, motivating the use of four instead of two overlapping offsets for subsequent real-data evaluations and as the repository default. The year-scale injection study instead uses two offsets to keep its computational cost tractable. Because the development background is limited, these results are used for comparative model, processing, and threshold selection rather than as an estimate of the final search background.

  
\subsubsection{Ablations summary}

The ablations favored a lightweight early-fusion TCN with shared per-window normalization. Normalization did not substantially change median performance, but it reduced seed-to-seed variation and improved training stability. For our BBH signals, a 1024-sample ($250\,\mathrm{ms}$) label performed better than the shorter 512-sample label. The preferred learning-rate scheduler used a patience of two epochs and an improvement threshold of $3\times10^{-4}$, indicating that relatively prompt learning-rate reductions were more effective than waiting longer for training to recover.

An $8\,\mathrm{s}$ whitening context was best among the tested $4$, $8$, $16$, and $32\,\mathrm{s}$ settings. Whitening appears to have an intermediate optimum: the context must be long enough to estimate the local noise spectrum reliably, but not so long that it becomes less representative of the immediate data. Longer contexts also impose unnecessary computational cost. 

Stricter glitch rejection was not beneficial; retaining a broad range of transient noise features allowed the model to learn realistic nonstationary backgrounds rather than training only on unusually clean data.

Among the trigger schemes, \texttt{smoothed\_max} gave the best overall performance across models, although \texttt{peak\_width} and \texttt{full\_peak} followed closely. The latter two performed nearly identically, indicating that the additional body-cut criterion in \texttt{full\_peak} provided little benefit. Peak-width settings were most effective when kept near the characteristic width used during training. These results favor a relatively simple trigger construction and suggest that additional post-processing complexity does not necessarily improve performance.

For final inference, we retain the \texttt{smoothed\_max} trigger with low- and high-SNR ranges of $[3,10]$ and $[10,25]$, with $p_{\mathrm{high}}$ decreasing from 0.90 to 0.25 and a probability of drawing noise-only samples of 0.65. Overall, SNR ablations indicate that the training distribution should extend well below the nominal detection threshold, but should not be concentrated too narrowly at the lowest SNRs. Curricula that combine very low-SNR examples with either a broader higher-SNR component or a larger fraction of noise-only samples produce the most robust sensitivity across changing ASD mismatch conditions. By contrast, curricula dominated by higher-SNR signals consistently underperform, while narrowly low-SNR curricula can lose robustness unless the training balance is adjusted.

The exact configuration and operating point used in the subsequent study on real O3b data, described in Sec.~\ref{methods:realdata}, follows directly from our ablation experiments, with two principal modifications. First, in line with preceding work~\cite{Nousi_2023}, we substantially extend the training-noise data set from the 17.1 hours used in the ablations to a livetime of over 12 days. Second, motivated by the observed sensitivity to window alignment, we increase from two to four inference offsets for cataloged-signal recovery and real-noise evaluation. The year-scale injection study retains two offsets for computational tractability.

\subsection{Benchmark on O3b data}

\subsubsection{Recovery of cataloged signals}
\label{sec:real_recovery}

We first evaluate four independently trained instances of the model on the 13 confident GWTC-3 events occurring within the held-out O3b evaluation interval. The training configuration, inference procedure, and \texttt{smoothed\_max} threshold of 0.9999 were fixed from the O3a development study before examining any O3b data, as described in Sec.~\ref{methods:realdata}. No data from the final January 28--February 29, 2020 evaluation interval was used for training, selection, or threshold tuning.

At this operating point, the four training seeds yield outcomes of 8/0, 9/0, 9/1, and 9/0 recovered events/false-positive triggers, respectively, over the 22.7 days of coincident O3b background livetime. With zero observed triggers over 22.7 days, the empirical false-alarm rate is unresolved below one trigger per analyzed livetime, corresponding to approximately 16 false positives per year. Under a stationary Poisson-process assumption, zero observed triggers over 22.7 days corresponds to a one-sided 95\% upper limit of approximately 48 false positives per year. Due to the short background, we interpret this evaluation as validation of the learned detector and trigger pipeline, rather than as a complete astrophysical search.

Eight of the 13 cataloged events are recovered by all four models, while GW200216\_220804 is additionally recovered by three of the four. Three of the models yield zero observed false positives, while one of them returns one trigger that does not correspond to any catalog events. Table~\ref{tab:o3b_catalog_events} summarizes the consistency of the catalog-event recoveries across seeds.

\begin{table*}[!h]
    \centering
    \small
    \setlength{\tabcolsep}{4pt}
    \renewcommand{\arraystretch}{1.1}
    \begin{tabular}{clcccc}
        \toprule
        & Event
        & $m_1^{\mathrm{source}}\,[M_\odot]$
        & $m_2^{\mathrm{source}}\,[M_\odot]$
        & $d_L\,[\mathrm{Mpc}]$
        & SNR \\
        \midrule

        \multirow{8}{*}{%
            {\scriptsize\shortstack[c]{Recovered\\by 4/4\\models}}}
        & GW200128\_022011
        & $42.2^{+11.6}_{-8.1}$
        & $32.6^{+9.5}_{-9.2}$
        & $3400^{+2100}_{-1800}$
        & $10.6^{+0.3}_{-0.4}$ \\

        & GW200129\_065458
        & $34.5^{+9.9}_{-3.1}$
        & $29.0^{+3.3}_{-9.3}$
        & $890^{+260}_{-370}$
        & $26.8^{+0.2}_{-0.2}$ \\

        & GW200208\_130117
        & $37.7^{+9.3}_{-6.2}$
        & $27.4^{+6.3}_{-7.3}$
        & $2230^{+1020}_{-850}$
        & $10.8^{+0.3}_{-0.4}$ \\

        & GW200209\_085452
        & $35.6^{+10.5}_{-6.8}$
        & $27.1^{+7.8}_{-7.8}$
        & $3400^{+1900}_{-1800}$
        & $9.6^{+0.4}_{-0.5}$ \\

        & GW200219\_094415
        & $37.5^{+10.1}_{-6.9}$
        & $27.9^{+7.4}_{-8.4}$
        & $3400^{+1700}_{-1500}$
        & $10.7^{+0.3}_{-0.5}$ \\

        & GW200220\_124850
        & $38.9^{+14.1}_{-8.6}$
        & $27.9^{+9.2}_{-9.0}$
        & $4000^{+2800}_{-2200}$
        & $8.5^{+0.3}_{-0.5}$ \\

        & GW200224\_222234
        & $40.0^{+6.7}_{-4.5}$
        & $32.7^{+4.8}_{-7.2}$
        & $1710^{+500}_{-650}$
        & $20.0^{+0.2}_{-0.2}$ \\

        & GW200225\_060421
        & $19.3^{+5.0}_{-3.0}$
        & $14.0^{+2.8}_{-3.5}$
        & $1150^{+510}_{-530}$
        & $12.5^{+0.3}_{-0.4}$ \\

        \midrule

        {\scriptsize
        \begin{tabular}[c]{@{}c@{}}
            Recovered\\
            by 3/4\\
            models
        \end{tabular}}
        & GW200216\_220804
        & $51^{+22}_{-13}$
        & $30^{+14}_{-16}$
        & $3800^{+3000}_{-2000}$
        & $8.1^{+0.4}_{-0.5}$ \\
        
        \midrule

        \multirow{4}{*}{%
            {\scriptsize\shortstack[c]{Recovered\\by 0/4\\models}}}
        & GW200202\_154313
        & $10.1^{+3.5}_{-1.4}$
        & $7.3^{+1.1}_{-1.7}$
        & $410^{+150}_{-160}$
        & $10.8^{+0.2}_{-0.4}$ \\

        & GW200208\_222617
        & $51^{+103}_{-30}$
        & $12.3^{+9.2}_{-5.5}$
        & $4100^{+4400}_{-2000}$
        & $7.4^{+1.4}_{-1.2}$ \\

        & GW200210\_092254
        & $24.1^{+7.5}_{-4.6}$
        & $2.83^{+0.47}_{-0.42}$
        & $940^{+430}_{-340}$
        & $8.4^{+0.5}_{-0.7}$ \\

        & GW200220\_061928
        & $87^{+40}_{-23}$
        & $61^{+26}_{-25}$
        & $6000^{+4800}_{-3100}$
        & $7.2^{+0.4}_{-0.7}$ \\

        \bottomrule
    \end{tabular}
    \caption{
        Source-frame component masses, luminosity distances, and network
        SNRs of the 13 confident GWTC-3 events in the evaluation interval,
        as reported by GWTC-3~\cite{KAGRA:2021vkt, GWOSC_GWTC3}. Events are grouped by the
        number of independently trained models that recover them at the
        predefined \texttt{smoothed\_max} threshold of 0.9999.
    }
    \label{tab:o3b_catalog_events}
\end{table*}

All triggers produced by the predefined \texttt{smoothed\_max} procedure are retained. We apply no post hoc trigger rejection, data-quality veto, spectrogram-based review, auxiliary-channel filtering, or coincidence across multiple independently trained models. Thus, the absence of observed false-positive triggers over the 22.7-day background for three of the four models follows directly from a single model using only a fixed-threshold trigger, without any subsequent candidate rejection.

\begin{figure*}[!h]
    \centering
    \includegraphics[width=\textwidth]
    {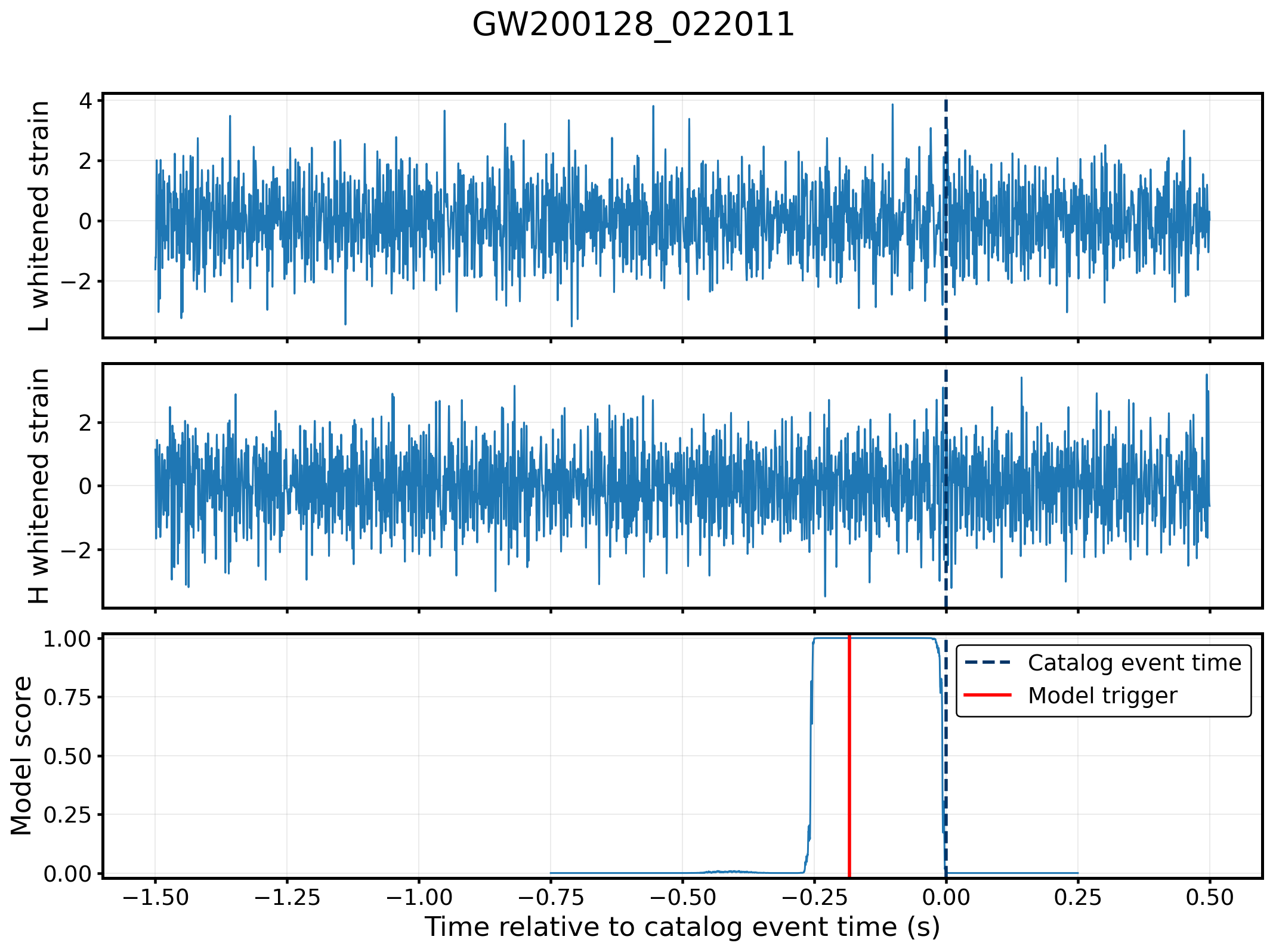}
    \caption{
    Representative recovery of the cataloged event GW200128\_022011 by the final early-fusion TCN. The upper panels show the whitened Livingston and Hanford strain, and the lower panel shows the raw per-timestep model score. The dashed blue line marks the catalog event time, while the solid red line marks the trigger time returned by the \texttt{smoothed\_max} procedure under a threshold of 0.9999.
    }
    \label{fig:o3b_score_example}
\end{figure*}

The four catalog events not recorded by any model are all modest-SNR detections, with network SNRs between \(7.2\) and \(10.8\), but span a broad range of source properties. Three are notable relative to the training population: GW200220\_061928 is the only one of all 13 catalog events whose median detector-frame masses lie outside the training mass distribution\footnote{Although Table~\ref{tab:o3b_catalog_events} reports source-frame component masses, consistency with the training distribution is assessed using detector-frame masses, related by \(m_i^{\mathrm{det}}=(1+z)m_i^{\mathrm{source}}\).}; GW200208\_222617 is an unusually asymmetric merger; while GW200210\_092254, also unusually asymmetric, contains a $2.83^{+0.47}_{-0.42},M_\odot$ secondary and may represent an NSBH rather than a BBH system.~\cite{GWOSC_GWTC3} As demonstrated in Figure~\ref{fig:signal_generalization}, a model trained on BBH merger signals may generalize poorly to NSBH systems. The latter two events have higher median mass ratios than any of the recovered signals, although both remain within the  mass bounds sampled for the training injection set. Thus, the unrecovered population includes one event outside the training mass distribution, one that may represent a source morphology not included in training, and one occupying an unusual but in-distribution region. GW200202\_154313, recovered by none of the four models, and GW200216\_220804, recovered by three of the four models, are comparatively typical in-distribution BBH mergers. 

These four unrecovered events are also challenging for established matched-filter searches. GW200202\_154313, GW200208\_222617, and GW200210\_092254 were not recovered significantly by either of the GstLAL analyses in Ref.~\cite{Joshi:2025zdu}, while GW200220\_061928 was identified significantly only by the PyCBC-BBH analysis among the five searches reported in GWTC-3~\cite{KAGRA:2021vkt}. This overlap suggests that the unrecovered events are not uniquely difficult for the neural pipeline.

Figure~\ref{fig:o3b_score_example} illustrates a characteristic output--returned by the model which recovered eight events with zero observed false positives. The score rises to near saturation over the 1024-sample (250ms) interval immediately preceding the catalog merger time, closely matching the width and placement of the positive label used during training. Most events recovered by this model exhibit the same broad, nearly saturated response. GW200202\_154313, GW200208\_222617, and GW200216\_220804 produce similarly elevated responses, but with small oscillations that keep the smoothed maximum below the threshold. By contrast, GW200210\_092254 and GW200220\_061928, the two potentially out of distribution events, produce substantially weaker responses. The strong but oscillatory responses of the remaining missed events suggest that a trigger definition designed to capture sustained score elevation more robustly could improve recovery.

\subsubsection{Comparison with prior machine-learning searches on O3 data}

\paragraph{Tian et al} The closest prior comparison on the same O3b noise period is the February 2020 benchmark of Tian et al.~\cite{Tian:2023vdc}. Unlike our models, their graph-based models were fine-tuned using noise drawn from the same February 2020 period. Individual models produced on the order of $10^2$--$10^3$ background triggers. Tian et al. therefore used coincidence across six independently trained models, after which 13 candidate triggers remained. Of these, seven  false positives were removed through data-quality checks and spectrogram-based review, leaving six candidates. 

In the present evaluation, the four independent early-fusion TCNs recover 8, 9, 9, and 9 of the 13 confident catalog events while producing 0, 0, 1, and 0 false-positive triggers, respectively, over 22.7 days of coincident background. The present pipeline therefore reaches a substantially cleaner single-model operating point on the February 2020 noise background than the earlier multi-model benchmark, without fine-tuning on the evaluation interval or relying on downstream coincidence or manual candidate-review stages.

\paragraph{AresGW} A broader comparison is provided by AresGW, which uses a deep 54-layer ResNet~\cite{Nousi_2023} and searched 6.7 months of O3 data and recovered 34 of 43 previously published events within its effective training range, corresponding to approximately $79\%$ recovery~\cite{Koloniari_2025}. The four models evaluated here recover $67$--$82\%$ of in-distribution catalog events.\footnote{The range reflects the recovery of eight or nine events across training seeds and the exclusion of one or two events outside the BBH training distribution.} These recovery fractions provide context for the catalog-event recovery achieved by the learned models, although AresGW evaluates a substantially longer observing interval and considers different in-distribution training populations. 

AresGW is a more complete search pipeline than ours: among other capabilities, it estimates the astrophysical probability $p_{astro}$ of its candidates and performs parameter estimation tests. These steps provide the statistical interpretation required for a production search and are not included in the present study. Its low-background search also includes multiple stages of candidate classification and noise rejection. Before trigger classification and the removal of known glitches, $\mathcal{O} (150)$ triggers remained in AresGW's initial candidate stream. Its pipeline reruns candidate intervals with several frequency filters, examines the density of nearby triggers, and rejects candidates whose times overlap glitches identified in an external Omicron and Gravity Spy catalog~\cite{ROBINET2020100620,Zevin_2017}. Remaining candidates are additionally inspected using time--frequency spectrograms, with further visually apparent glitches removed. 

At the single-model level, the present models produce zero or one false-positive trigger in 22.7 days using only a fixed threshold and predefined smoothing operation, without applying external glitch lookup, auxiliary veto, or candidate review. Together with the ablation result that stricter glitch removal during training did not improve convergence, these results indicate that the lightweight TCN maintains a low raw trigger count in the presence of the transient noise features encountered in the evaluation data.

The aim of this comparison is not to establish that the complete pipeline is better than AresGW, but to assess and improve robustness at the level of a single learned model. These results indicate that, at the single-model level, a substantially simpler learned detection and trigger stage may be able to achieve numerically similar catalog-event recovery while producing a cleaner candidate stream before downstream processing.

\subsubsection{Injection study}

\begin{figure*}[!h]
    \centering
    \includegraphics[width=0.9\textwidth]{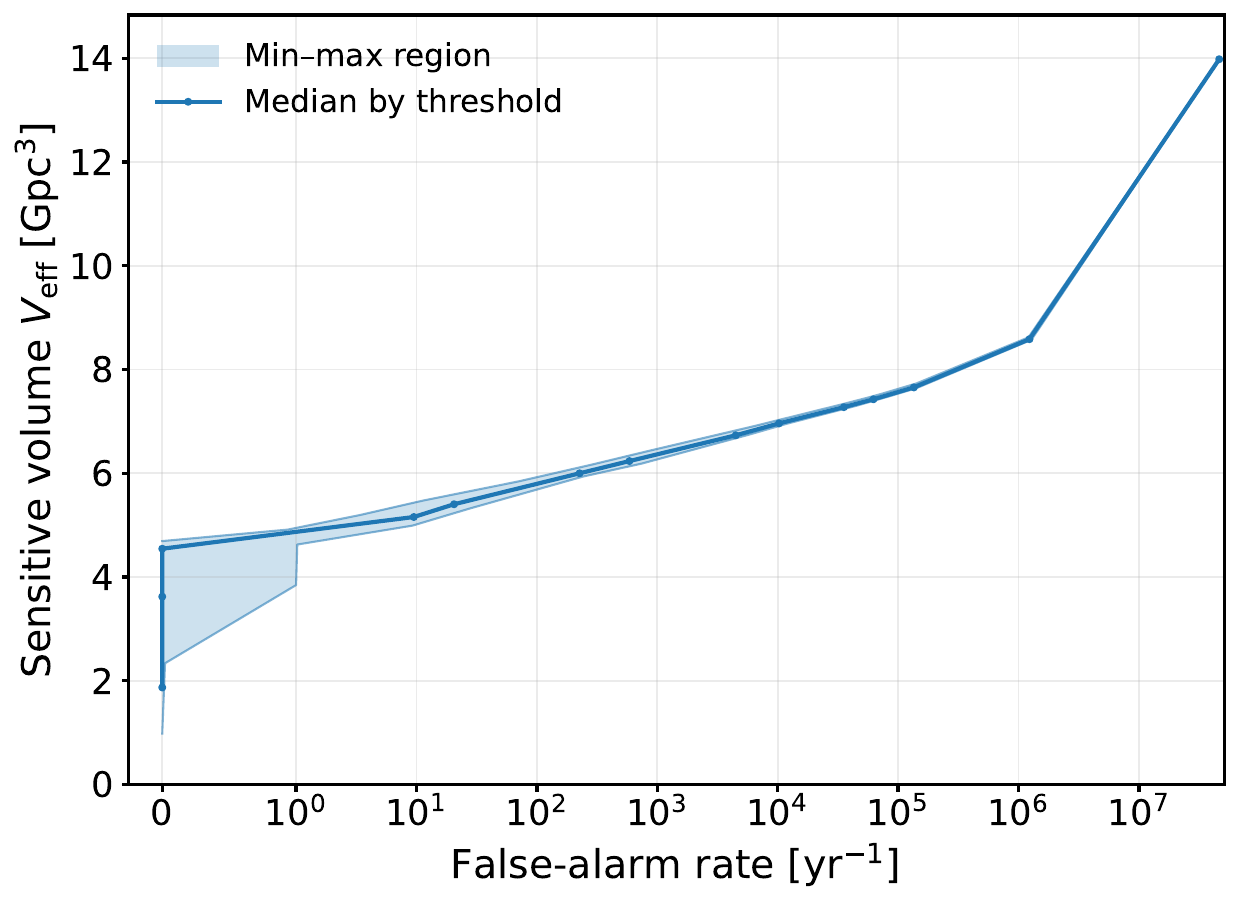}
    \caption{Sensitive volume as a function of false-alarm rate across four training seeds using \texttt{smoothed\_max}. The shaded band shows the min–max seed envelope, and the solid curve shows the median false-alarm rate and sensitive volume at each threshold. The false-alarm rates are empirical values obtained from the observed trigger counts over the 365-day background livetime.
    }
    \label{fig:o3b_injections}
\end{figure*}

The complementary injection study on the held-out O3b data is shown in Fig.~\ref{fig:o3b_injections}. For the injection study, we constructed 365 days of effective noise background and evaluated 10,000 signal injections. The corresponding sensitive-volume curves split into four total-mass bins are shown in~\ref{app:mass_binned_sensitive_volume}.

At the threshold of \(0.9999\) selected from the O3a development study, the four models achieve a median sensitive volume of \(5.40\,\mathrm{Gpc}^3\) at a median false-alarm rate of \(20.5\,\mathrm{yr}^{-1}\). We also examine two stricter thresholds as supplementary operating points. At \(0.99995\), the median sensitive volume is \(5.16\,\mathrm{Gpc}^3\) at \(9.51\,\mathrm{yr}^{-1}\) false alarms. At that point, the models recover 9, 7, 9, and 9 cataloged events with no false positives on the February 2020 data. At \(0.99999\), the median sensitive volume decreases to \(4.55\,\mathrm{Gpc}^3\). At that threshold, one model produces 1 false positive in its one-year background, while the other three produce none. At that operating point, the models recover 7, 5, 7, and 7 cataloged events with no false positives in the February 2020 dataset. For zero observed triggers over one year of background, the corresponding one-sided 95\% Poisson upper limit is a FAR of \(3.0\,\mathrm{yr}^{-1}\).

\section{Discussion}
\label{sec:discussion}

This study combines an end-to-end detection pipeline, a broad ablation of its architectural, training, preprocessing, and inference components, and a final single-model configuration selected for evaluation on held-out O3b data.

Across these experiments, the best configuration was not obtained by maximizing either model complexity or data cleanliness. Within the tested implementations and common training configuration, simple TCN architectures generally outperformed the HDCN-based, attention-based, and other more elaborate models, with the early-fusion TCN providing the best combination of sensitivity, stability, and architectural simplicity for the final analysis.\footnote{The original \textsc{AttenGW} cross-attention model nevertheless reduced the background relative to the graph-based architecture it was designed to replace and is therefore retained as a separately validated intermediate result, see ~\ref{app:attengw_validation}} This result should not be interpreted as a general limitation of these architecture families, since their performance may change with architecture-specific optimization and under different data realizations.

More broadly, performance was strongest when each component of the pipeline supplied enough information and difficulty to be useful without introducing unnecessary complexity. Shared per-window normalization reduced seed-to-seed variation, an intermediate whitening context outperformed both shorter and longer alternatives, and a longer label width was preferable to a shorter target. Stricter glitch rejection did not improve training, suggesting that retaining realistic nonstationary noise features is more useful than restricting training to unusually clean data. Similarly, the preferred SNR curriculum extends well below the eventual detection threshold while retaining a broad higher-SNR component, rather than concentrating training entirely on either easy or near-threshold examples. At inference time, the simple \texttt{smoothed\_max} trigger outperformed the more elaborate width-based alternatives overall.

The multi-seed and ASD-mismatch experiments provide complementary robustness tests. Shared normalization reduced seed-to-seed variation, which is particularly important because each reported search result is produced by a single model rather than by selecting or combining multiple networks. The ASD-mismatch partition shows how sensitivity changes as evaluation noise becomes less similar to the training distribution: performance generally declines with increasing mismatch, although the strength of the trend varies across models and configurations. While not a calibrated uncertainty estimate, ASD mismatch provides a useful model-agnostic diagnostic of domain shift that can be computed before interpreting detections.

The final O3b evaluation demonstrates the practical consequence of these choices. Four independently trained lightweight models recover 8, 9, 9, and 9 of the 13 confident GWTC-3 events in the held-out interval while producing 0, 0, 1, and 0 false-positive triggers, respectively, over 22.7 days of coincident Hanford--Livingston background livetime. All four recover the same eight events, with the three higher-performing models additionally recovering GW200216\_220804. The training data are drawn from the strictly preceding O3b period, but the architecture, training configuration, checkpoint-selection rule, inference procedure, and threshold are selected without using data from the January 28--February 29, 2020 evaluation interval. All candidates produced by the fixed \texttt{smoothed\_max} trigger are retained. The absence of observed triggers with three of the four models in this background interval therefore reflects the combination of a single learned detection statistic and a predefined, minimal trigger construction rather than a subsequent candidate-cleaning stage.

This distinction is important for both interpretability and reproducibility. When performance depends on post hoc vetoes, manual review, or coincidence across many independently trained networks, it becomes harder to determine whether the detector itself has learned a robust decision rule or whether errors are being removed downstream. Our results instead point toward a promising direction for machine-learning searches: models with low seed-to-seed variation and strong single-model performance that only minimal, predefined trigger construction is required. Reducing reliance on non-ML rescue stages should make such systems easier to validate, reproduce, and eventually deploy in broader searches.

Future work should broaden validation across observing periods and beyond the few months of O3a and O3b data used here; detector conditions; signal populations beyond this small set of compact-binary populations; and input channels. The TCN early-fusion architecture can accommodate additional channels through a simple increase in the configured input dimensionality, although this extension remains untested. The observed sensitivity to signal class and SNR curriculum motivates efforts to improve robustness across population choices or, alternatively, to exploit such specialization through ensembles or multi-model classification. Reducing the remaining seed-to-seed variation is another important direction. ASD mismatch provides a useful first diagnostic of distribution shift, but additional measures would characterize changes between training and evaluation data more completely. Importantly, calibrating the output scores as calibrated probabilities~\cite{guo2017calibrationmodernneuralnetworks} could support more principled trigger selection and clearer interpretation. Future work could reserve a chronologically intermediate noise interval for checkpoint and operating-point selection, particularly given the observed seed-to-seed variation, or explore periodic recalibration and retraining as detector conditions evolve.

The released software~\cite{AttenGWRepo} includes the full sequence from GWOSC data acquisition and PSD estimation through waveform injection, training, checkpoint selection, continuous inference, trigger construction, and evaluation. The small model size and substantially faster-than-real-time inference make the pipeline practical for systematic experiments on shared computing resources.


\subsection{Software Availability}
The software used to download and preprocess GWOSC strain, construct training examples from user-provided waveform injections, train the neural-network architectures considered in this study, compute the ASD-mismatch diagnostic, and perform checkpoint-based inference and trigger construction is publicly available at \url{https://github.com/victoria-tiki/AttenGW}. The repository also provides example signal and noise data for testing, a handbook for optimizing models, and selected O3b model checkpoints together with their saved training configurations. The repository retains the name \textsc{AttenGW} for historical reasons, reflecting the cross-attention model for which it was originally developed; the current code supports the broader, architecture-independent pipeline used in this study.

\section*{Acknowledgements}

We thank Roland Haas, Michael Pürrer, and Minyang Tian for their assistance with data generation. We also thank Prayush Kumar, Vitor Ramos, and Arjun Chainani for insightful discussions that helped shape this work.

We gratefully acknowledge
support from National Science Foundation awards OAC-1931561 and OAC-2209892.
This research has made use of data
obtained from the Gravitational Wave Open Science
Center (https://gwosc.org), a service of the LIGO Scientific Collaboration, the Virgo Collaboration, and KAGRA. This work was partially supported by the U.S.
Department of Energy under Contract No. DE-AC02-06CH11357, including funding from the Office of Advanced Scientific Computing Research (ASCR)’s
Diaspora project and the Laboratory Directed Research and
Development program. This research used both the
DeltaAI advanced computing and data resource, which
is supported by the National Science Foundation (award
OAC 2320345) and the State of Illinois, and the Delta
advanced computing and data resource which is supported by the National Science Foundation (award OAC
2005572) and the State of Illinois. Delta and DeltaAI
are joint efforts of the University of Illinois Urbana-Champaign and its National Center for Supercomputing
Applications.



\bibliographystyle{elsarticle-num}
\bibliography{refs}

@ARTICLE{2019ApJ...882L..24A,
author = {B. P. Abbott and others},
collaboration = {(LIGO Scientific Collaboration and Virgo Collaboration)},
        title = "{Binary Black Hole Population Properties Inferred from the First and Second Observing Runs of Advanced LIGO and Advanced Virgo}",
      journal = {Astrophys. J. Lett.},
         year = 2019,
        month = sep,
       volume = {882},
       number = {2},
          eid = {L24},
        pages = {L24},
          doi = {10.3847/2041-8213/ab3800},
archivePrefix = {arXiv},
       eprint = {1811.12940},
 primaryClass = {astro-ph.HE},
       adsurl = {https://ui.adsabs.harvard.edu/abs/2019ApJ...882L..24A}
}

@article{Joshi:2025zdu,
    author = "Joshi, Prathamesh and others",
    title = "{How Many Times Should We Matched Filter Gravitational Wave Data? A Comparison of GstLAL's Online and Offline Performance}",
    eprint = "2505.23959",
    archivePrefix = "arXiv",
    primaryClass = "gr-qc",
    month = "5",
    year = "2025"
}

@article{ROBINET2020100620,
title = {Omicron: A tool to characterize transient noise in gravitational-wave detectors},
journal = {SoftwareX},
volume = {12},
pages = {100620},
year = {2020},
issn = {2352-7110},
doi = {10.1016/j.softx.2020.100620},
url = {https://www.sciencedirect.com/science/article/pii/S2352711020303332},
author = {Florent Robinet and Nicolas Arnaud and Nicolas Leroy and Andrew Lundgren and Duncan Macleod and Jessica McIver},
}

@article{MORENOTORRES2012521,
title = {A unifying view on dataset shift in classification},
journal = {Pattern Recognition},
volume = {45},
number = {1},
pages = {521-530},
year = {2012},
issn = {0031-3203},
doi = {10.1016/j.patcog.2011.06.019},
url = {https://www.sciencedirect.com/science/article/pii/S0031320311002901},
author = {Jose G. Moreno-Torres and Troy Raeder and Rocío Alaiz-Rodríguez and Nitesh V. Chawla and Francisco Herrera},
}

@article{poissongarwood,
  author  = {Garwood, F.},
  title   = {Fiducial Limits for the Poisson Distribution},
  journal = {Biometrika},
  volume  = {28},
  number  = {3--4},
  pages   = {437--442},
  year    = {1936},
  month   = dec,
  doi     = {10.1093/biomet/28.3-4.437},
  url     = {https://doi.org/10.1093/biomet/28.3-4.437}
}

@misc{2018arXiv181011953R,
  author        = {Rabanser, Stephan and G{\"u}nnemann, Stephan
                   and Lipton, Zachary C.},
  title         = {Failing Loudly: An Empirical Study of Methods
                   for Detecting Dataset Shift},
  year          = {2018},
  eprint        = {1810.11953},
  archivePrefix = {arXiv},
  primaryClass  = {stat.ML},
  doi           = {10.48550/arXiv.1810.11953}
}

@article{Zevin_2017,
doi = {10.1088/1361-6382/aa5cea},
url = {https://doi.org/10.1088/1361-6382/aa5cea},
year = {2017},
month = {feb},
publisher = {IOP Publishing},
volume = {34},
number = {6},
pages = {064003},
author = {Zevin, M and Coughlin, S and Bahaadini, S and Besler, E and Rohani, N and Allen, S and Cabero, M and Crowston, K and Katsaggelos, A K and Larson, S L and Lee, T K and Lintott, C and Littenberg, T B and Lundgren, A and Østerlund, C and Smith, J R and Trouille, L and Kalogera, V},
title = {Gravity Spy: integrating advanced LIGO detector characterization, machine learning, and citizen science},
journal = {Classical and Quantum Gravity},
}

@article{LIGO:2021ppb,
    author = "Davis, Derek and others",
    collaboration = "LIGO",
    title = "{LIGO detector characterization in the second and third observing runs}",
    eprint = "2101.11673",
    archivePrefix = "arXiv",
    primaryClass = "astro-ph.IM",
    reportNumber = "P2000495",
    doi = "10.1088/1361-6382/abfd85",
    journal = "Class. Quant. Grav.",
    volume = "38",
    number = "13",
    pages = "135014",
    year = "2021"
}

@misc{Falcon2019Lightning,
  author       = {Falcon, William and {{The PyTorch Lightning team}}},
  title        = {{PyTorch Lightning}},
  year         = {2019},
  month        = mar,
  note         = {Computer software},
  doi          = {10.5281/zenodo.3530844},
  url          = {https://www.pytorchlightning.ai}
}

@article{LIGOScientific:2019gag,
    author = "Abbott, B. P. and others",
    collaboration = "LIGO Scientific, Virgo",
    title = "{Low-latency Gravitational-wave Alerts for Multimessenger Astronomy during the Second Advanced LIGO and Virgo Observing Run}",
    eprint = "1901.03310",
    archivePrefix = "arXiv",
    primaryClass = "astro-ph.HE",
    reportNumber = "LIGO-P1800255",
    doi = "10.3847/1538-4357/ab0e8f",
    journal = "Astrophys. J.",
    volume = "875",
    number = "2",
    pages = "161",
    year = "2019"
}

@article{Koloniari_2025,
   title={New gravitational wave discoveries enabled by machine learning},
   volume={6},
   ISSN={2632-2153},
   url={http://dx.doi.org/10.1088/2632-2153/adb5ed},
   DOI={10.1088/2632-2153/adb5ed},
   number={1},
   journal={Machine Learning: Science and Technology},
   publisher={IOP Publishing},
   author={Koloniari, Alexandra E and Koursoumpa, Evdokia C and Nousi, Paraskevi and Lampropoulos, Paraskevas and Passalis, Nikolaos and Tefas, Anastasios and Stergioulas, Nikolaos},
   year={2025},
   month=Feb, pages={015054} }

@article{KAGRA:2023pio,
    author = "Abbott, R. and others",
    collaboration = "KAGRA, VIRGO, LIGO Scientific",
    title = "{Open Data from the Third Observing Run of LIGO, Virgo, KAGRA, and GEO}",
    eprint = "2302.03676",
    archivePrefix = "arXiv",
    primaryClass = "gr-qc",
    reportNumber = "LIGO-P2200316",
    doi = "10.3847/1538-4365/acdc9f",
    journal = "Astrophys. J. Suppl.",
    volume = "267",
    number = "2",
    pages = "29",
    year = "2023"
}

@article{scipyfindpeaks,
  author  = {Virtanen, Pauli and Gommers, Ralf and Oliphant, Travis E. and others},
    year = {2020},
    month = {02},
    pages = {261--272},
    title = {{SciPy} 1.0: Fundamental Algorithms for Scientific Computing in {Python}},
    volume = {17},
    journal = {Nature Methods},
    doi = {10.1038/s41592-019-0686-2}
}

@article{Schaefer2023MLGWSC,
    author = {Sch{\"a}fer, Marlin B. and others},
    title = "{First machine learning gravitational-wave search mock data challenge}",
    eprint = "2209.11146",
    archivePrefix = "arXiv",
    primaryClass = "astro-ph.IM",
    doi = "10.1103/PhysRevD.107.023021",
    journal = "Phys. Rev. D",
    volume = "107",
    number = "2",
    pages = "023021",
    year = "2023"
}

@article{Glanzer_2023,
   title={Data quality up to the third observing run of advanced LIGO: Gravity Spy glitch classifications},
   volume={40},
   ISSN={1361-6382},
   url={http://dx.doi.org/10.1088/1361-6382/acb633},
   DOI={10.1088/1361-6382/acb633},
   number={6},
   journal={Classical and Quantum Gravity},
   publisher={IOP Publishing},
   author={Glanzer, J and Banagiri, S and Coughlin, S B and Soni, S and Zevin, M and Berry, C P L and Patane, O and Bahaadini, S and Rohani, N and Crowston, K and Kalogera, V and Østerlund, C and Trouille, L and Katsaggelos, A},
   year={2023},
   month=Feb, pages={065004} }

@article{Gebhard2019,
  title = {Convolutional neural networks: A magic bullet for gravitational-wave detection?},
  author = {Gebhard, Timothy D. and Kilbertus, Niki and Harry, Ian and Sch\"olkopf, Bernhard},
  journal = {Phys. Rev. D},
  volume = {100},
  issue = {6},
  pages = {063015},
  numpages = {19},
  year = {2019},
  month = {Sep},
  publisher = {American Physical Society},
  doi = {10.1103/PhysRevD.100.063015},
  url = {https://link.aps.org/doi/10.1103/PhysRevD.100.063015}
}

@article{Gabbard2018,
  title = {Matching Matched Filtering with Deep Networks for Gravitational-Wave Astronomy},
  author = {Gabbard, Hunter and Williams, Michael and Hayes, Fergus and Messenger, Chris},
  journal = {Phys. Rev. Lett.},
  volume = {120},
  issue = {14},
  pages = {141103},
  numpages = {6},
  year = {2018},
  month = {Apr},
  publisher = {American Physical Society},
  doi = {10.1103/PhysRevLett.120.141103},
  url = {https://link.aps.org/doi/10.1103/PhysRevLett.120.141103}
}

@misc{paszke2019pytorchimperativestylehighperformance,
      title={{PyTorch}: An Imperative Style, High-Performance Deep Learning Library},
      author={Adam Paszke and Sam Gross and Francisco Massa and Adam Lerer and James Bradbury and Gregory Chanan and Trevor Killeen and Zeming Lin and Natalia Gimelshein and Luca Antiga and Alban Desmaison and Andreas Köpf and Edward Yang and Zach DeVito and Martin Raison and Alykhan Tejani and Sasank Chilamkurthy and Benoit Steiner and Lu Fang and Junjie Bai and Soumith Chintala},
      year={2019},
      eprint={1912.01703},
      archivePrefix={arXiv},
      primaryClass={cs.LG},
      url={https://arxiv.org/abs/1912.01703}, 
}

@article{PhysRevD.60.022002,
  title = {Matched filtering of gravitational waves from inspiraling compact binaries: Computational cost and template placement},
  author = {Owen, Benjamin J. and Sathyaprakash, B. S.},
  journal = {Phys. Rev. D},
  volume = {60},
  issue = {2},
  pages = {022002},
  numpages = {12},
  year = {1999},
  month = {Jun},
  publisher = {American Physical Society},
  doi = {10.1103/PhysRevD.60.022002},
  url = {https://link.aps.org/doi/10.1103/PhysRevD.60.022002}
}

@misc{guo2017calibrationmodernneuralnetworks,
      title={On Calibration of Modern Neural Networks}, 
      author={Chuan Guo and Geoff Pleiss and Yu Sun and Kilian Q. Weinberger},
      year={2017},
      eprint={1706.04599},
      archivePrefix={arXiv},
      primaryClass={cs.LG},
      url={https://arxiv.org/abs/1706.04599}, 
}

@article{KAGRA:2021vkt,
    author = "Abbott, R. and others",
    collaboration = "KAGRA, VIRGO, LIGO Scientific",
    title = "{GWTC-3: Compact Binary Coalescences Observed by LIGO and Virgo during the Second Part of the Third Observing Run}",
    eprint = "2111.03606",
    archivePrefix = "arXiv",
    primaryClass = "gr-qc",
    reportNumber = "LIGO-P2000318",
    doi = "10.1103/PhysRevX.13.041039",
    journal = "Phys. Rev. X",
    volume = "13",
    number = "4",
    pages = "041039",
    year = "2023"
}

@inproceedings{Louradour_curriculum,
author = {Bengio, Yoshua and Louradour, J{\'e}r{\^o}me and Collobert, Ronan and Weston, Jason},
title = {Curriculum learning},
year = {2009},
isbn = {9781605585161},
publisher = {Association for Computing Machinery},
address = {New York, NY, USA},
url = {https://doi.org/10.1145/1553374.1553380},
doi = {10.1145/1553374.1553380},
booktitle = {Proceedings of the 26th Annual International Conference on Machine Learning},
pages = {41–48},
numpages = {8},
location = {Montreal, Quebec, Canada},
series = {ICML '09}
}

@article{W_s_2009,
   title={On the background estimation by time slides in a network of gravitational wave detectors},
   volume={27},
   ISSN={1361-6382},
   url={http://dx.doi.org/10.1088/0264-9381/27/1/015005},
   DOI={10.1088/0264-9381/27/1/015005},
   number={1},
   journal={Classical and Quantum Gravity},
   publisher={IOP Publishing},
   author={Wąs, Michał and Bizouard, Marie-Anne and Brisson, Violette and Cavalier, Fabien and Davier, Michel and Hello, Patrice and Leroy, Nicolas and Robinet, Florent and Vavoulidis, Miltiadis},
   year={2009},
   month=Dec, pages={015005} }

@misc{thompson2020modelinggravitationalwavesignature,
      title={Modeling the Gravitational-Wave Signature of Neutron Star--Black Hole Coalescences: {PhenomNSBH}}, 
      author={Jonathan E. Thompson and Edward Fauchon-Jones and Sebastian Khan and Elisa Nitoglia and Francesco Pannarale and Tim Dietrich and Mark Hannam},
      year={2020},
      eprint={2002.08383},
      archivePrefix={arXiv},
      primaryClass={gr-qc},
      doi={10.1103/PhysRevD.101.124059},
      url={https://arxiv.org/abs/2002.08383}, 
}

@article{Pratten:2020ceb,
    author = "Pratten, Geraint and others",
    title = "{Computationally efficient models for the dominant and subdominant harmonic modes of precessing binary black holes}",
    eprint = "2004.06503",
    archivePrefix = "arXiv",
    primaryClass = "gr-qc",
    doi = "10.1103/PhysRevD.103.104056",
    journal = "Phys. Rev. D",
    volume = "103",
    number = "10",
    pages = "104056",
    year = "2021"
}

@article{PhysRevD.102.064002,
  title = {Multimode frequency-domain model for the gravitational wave signal from nonprecessing black-hole binaries},
  author = {Garc\'{\i}a-Quir\'os, Cecilio and Colleoni, Marta and Husa, Sascha and Estell\'es, H\'ector and Pratten, Geraint and Ramos-Buades, Antoni and Mateu-Lucena, Maite and Jaume, Rafel},
  journal = {Phys. Rev. D},
  volume = {102},
  issue = {6},
  pages = {064002},
  numpages = {33},
  year = {2020},
  month = {Sep},
  publisher = {American Physical Society},
  doi = {10.1103/PhysRevD.102.064002},
  url = {https://link.aps.org/doi/10.1103/PhysRevD.102.064002}
}

@article{Welch1967TheUO,
  title={The use of fast Fourier transform for the estimation of power spectra: A method based on time averaging over short, modified periodograms},
  author={Peter D. Welch},
  journal={IEEE Transactions on Audio and Electroacoustics},
  year={1967},
  volume={15},
  pages={70-73},
  url={https://api.semanticscholar.org/CorpusID:13900622}
}

@article{Nousi_2023,
   title={Deep residual networks for gravitational wave detection},
   volume={108},
   ISSN={2470-0029},
   url={http://dx.doi.org/10.1103/PhysRevD.108.024022},
   DOI={10.1103/physrevd.108.024022},
   number={2},
   journal={Physical Review D},
   publisher={American Physical Society (APS)},
   author={Nousi, Paraskevi and Koloniari, Alexandra E. and Passalis, Nikolaos and Iosif, Panagiotis and Stergioulas, Nikolaos and Tefas, Anastasios},
   year={2023},
   month=July }

@article{bai2018empirical,
  author = {Bai, Shaojie and Kolter, J. Zico and Koltun, Vladlen},
  title = {An Empirical Evaluation of Generic Convolutional and Recurrent Networks for Sequence Modeling},
  year = {2018},
  eprint = {1803.01271},
  archivePrefix = {arXiv},
  primaryClass = {cs.LG}
}

@inproceedings{he2016deep,
  author    = {He, Kaiming and Zhang, Xiangyu and Ren, Shaoqing and Sun, Jian},
  title     = {Deep Residual Learning for Image Recognition},
  booktitle = {Proceedings of the IEEE Conference on Computer Vision and
               Pattern Recognition},
  pages     = {770--778},
  year      = {2016},
  doi       = {10.1109/CVPR.2016.90}
}

@article{Marx2025Aframe,
  author  = {Marx, Ethan and Benoit, William and Gunny, Alec and others},
  title   = {Machine-learning pipeline for real-time detection of
             gravitational waves from compact binary coalescences},
  journal = {Physical Review D},
  volume  = {111},
  number  = {4},
  pages   = {042010},
  year    = {2025},
  doi     = {10.1103/PhysRevD.111.042010},
  eprint  = {2403.18661},
  archivePrefix = {arXiv},
  primaryClass  = {gr-qc}
}

@misc{GWOSC_GWTC3,
    author       = {{LIGO Scientific Collaboration} and
                    {Virgo Collaboration} and
                    {KAGRA Collaboration}},
    title        = {{GWTC-3 Data Release}},
    year         = {2023},
    howpublished = {\url{https://gwosc.org/GWTC-3/}},
    note         = {Gravitational Wave Open Science Center,
                    accessed 2026-08-18}
}

@article{ba2016layer,
    author = {Ba, Jimmy Lei and Kiros, Jamie Ryan and Hinton, Geoffrey E.},
    title = {Layer Normalization},
    year = {2016},
    eprint = {1607.06450},
    archivePrefix = {arXiv},
    primaryClass = {stat.ML}
}

@inproceedings{zhu2024visionmamba,
  title     = {Vision Mamba: Efficient Visual Representation Learning with Bidirectional State Space Model},
  author    = {Zhu, Lianghui and Liao, Bencheng and Zhang, Qian and Wang, Xinlong and Liu, Wenyu and Wang, Xinggang},
  booktitle = {Proceedings of the 41st International Conference on Machine Learning},
  pages     = {62429--62442},
  year      = {2024},
  volume    = {235},
  series    = {Proceedings of Machine Learning Research},
  publisher = {PMLR}
}

@inproceedings{conneau2017supervised,
    title = {Supervised Learning of Universal Sentence Representations from Natural Language Inference Data},
    author = {Conneau, Alexis and Kiela, Douwe and Schwenk, Holger and Barrault, Lo{\"i}c and Bordes, Antoine},
    booktitle = {Proceedings of the 2017 Conference on Empirical Methods in Natural Language Processing},
    year = {2017},
    pages = {670--680},
    publisher = {Association for Computational Linguistics},
    doi = {10.18653/v1/D17-1070}
}

@inproceedings{lin2017feature,
  author    = {Lin, Tsung-Yi and Doll{\'a}r, Piotr and Girshick, Ross
               and He, Kaiming and Hariharan, Bharath and Belongie, Serge},
  title     = {Feature Pyramid Networks for Object Detection},
  booktitle = {Proceedings of the IEEE Conference on Computer Vision and
               Pattern Recognition},
  pages     = {2117--2125},
  year      = {2017}
}

@inproceedings{nie2023timeseries,
  author    = {Nie, Yuqi and Nguyen, Nam H. and Sinthong, Phanwadee
               and Kalagnanam, Jayant},
  title     = {A Time Series is Worth 64 Words: Long-term Forecasting
               with Transformers},
  booktitle = {The Eleventh International Conference on Learning Representations},
  year      = {2023}
}

@inproceedings{gu2024mamba,
  author    = {Gu, Albert and Dao, Tri},
  title     = {Mamba: Linear-Time Sequence Modeling with Selective State Spaces},
  booktitle = {The First Conference on Language Modeling},
  year      = {2024}
}

@article{oord2016wavenet,
  author = {van den Oord, Aaron and Dieleman, Sander and Zen, Heiga and
            Simonyan, Karen and Vinyals, Oriol and Graves, Alex and
            Kalchbrenner, Nal and Senior, Andrew and Kavukcuoglu, Koray},
  title = {{WaveNet}: A Generative Model for Raw Audio},
  year = {2016},
  eprint = {1609.03499},
  archivePrefix = {arXiv}
}

@inproceedings{vaswani2017attention,
  author    = {Vaswani, Ashish and Shazeer, Noam and Parmar, Niki
               and Uszkoreit, Jakob and Jones, Llion and Gomez, Aidan N.
               and Kaiser, Lukasz and Polosukhin, Illia},
  title     = {Attention Is All You Need},
  booktitle = {Advances in Neural Information Processing Systems},
  volume    = {30},
  year      = {2017}
}

@article{PhysRevLett.123.011102,
  title = {Tests of General Relativity with {GW170817}},
author = {B. P. Abbott and others},
collaboration = {LIGO Scientific Collaboration and Virgo Collaboration},
  journal = {Phys. Rev. Lett.},
  volume = {123},
  issue = {1},
  pages = {011102},
  numpages = {15},
  year = {2019},
  month = {Jul},
  publisher = {American Physical Society},
  doi = {10.1103/PhysRevLett.123.011102},
  url = {https://link.aps.org/doi/10.1103/PhysRevLett.123.011102}
}

@article{PhysRevLett.123.111102,
  title = {Testing the No-Hair Theorem with GW150914},
  author = {Isi, Maximiliano and Giesler, Matthew and Farr, Will M. and Scheel, Mark A. and Teukolsky, Saul A.},
  journal = {Phys. Rev. Lett.},
  volume = {123},
  issue = {11},
  pages = {111102},
  numpages = {6},
  year = {2019},
  month = {Sep},
  publisher = {American Physical Society},
  doi = {10.1103/PhysRevLett.123.111102},
  url = {https://link.aps.org/doi/10.1103/PhysRevLett.123.111102}
}

@article{Zevin_2020,
   title = {You Can’t Always Get What You Want: The Impact of Prior Assumptions on Interpreting {GW190412}},
   volume={899},
   ISSN={2041-8213},
   url={http://dx.doi.org/10.3847/2041-8213/aba8ef},
   DOI={10.3847/2041-8213/aba8ef},
   number={1},
   journal={The Astrophysical Journal Letters},
   publisher={American Astronomical Society},
   author={Zevin, Michael and Berry, Christopher P. L. and Coughlin, Scott and Chatziioannou, Katerina and Vitale, Salvatore},
   year={2020},
   month=aug, pages={L17} }

@article{Usman:2015kfa,
    author = "Usman, Samantha A. and others",
    title = "{The PyCBC search for gravitational waves from compact binary coalescence}",
    eprint = "1508.02357",
    archivePrefix = "arXiv",
    primaryClass = "gr-qc",
    reportNumber = "LIGO-P1500086",
    doi = "10.1088/0264-9381/33/21/215004",
    journal = "Class. Quant. Grav.",
    volume = "33",
    number = "21",
    pages = "215004",
    year = "2016"
}

@article{Messick_2017,
  author  = {Messick, Cody and others},
  title   = {Analysis Framework for the Prompt Discovery of Compact
             Binary Mergers in Gravitational-Wave Data},
  journal = {Physical Review D},
  volume  = {95},
  number  = {4},
  pages   = {042001},
  year    = {2017},
  month   = feb,
  doi     = {10.1103/PhysRevD.95.042001}
}

@article{George_2018,
   title={Deep neural networks to enable real-time multimessenger astrophysics},
   volume={97},
   ISSN={2470-0029},
   url={http://dx.doi.org/10.1103/PhysRevD.97.044039},
   DOI={10.1103/physrevd.97.044039},
   number={4},
   journal={Physical Review D},
   publisher={American Physical Society (APS)},
   author={George, Daniel and Huerta, E. A.},
   year={2018},
   month=feb }

@article{PhysRevD.101.104003,
  title = {Gravitational-Wave Signal Recognition of {LIGO} Data by Deep Learning},
  author = {Wang, He and Wu, Shichao and Cao, Zhoujian and Liu, Xiaolin and Zhu, Jian-Yang},
  journal = {Phys. Rev. D},
  volume = {101},
  issue = {10},
  pages = {104003},
  numpages = {10},
  year = {2020},
  month = {May},
  publisher = {American Physical Society},
  doi = {10.1103/PhysRevD.101.104003},
  url = {https://link.aps.org/doi/10.1103/PhysRevD.101.104003}
}

@article{QIU2023137850,
title = {Deep learning detection and classification of gravitational waves from neutron star-black hole mergers},
journal = {Physics Letters B},
volume = {840},
pages = {137850},
year = {2023},
issn = {0370-2693},
doi = {10.1016/j.physletb.2023.137850},
url = {https://www.sciencedirect.com/science/article/pii/S0370269323001843},
author = {Richard Qiu and Plamen G. Krastev and Kiranjyot Gill and Edo Berger},
}

@article{Tian:2023vdc,
    author = "Tian, Minyang and Huerta, E. A. and Zheng, Huihuo and Kumar, Prayush",
    title = "{Physics-inspired spatiotemporal-graph AI ensemble for the detection of higher order wave mode signals of spinning binary black hole mergers}",
    eprint = "2306.15728",
    archivePrefix = "arXiv",
    primaryClass = "astro-ph.IM",
    doi = "10.1088/2632-2153/ad4c37",
    journal = "Mach. Learn. Sci. Tech.",
    volume = "5",
    number = "2",
    pages = "025056",
    year = "2024"
}

@misc{mobilia2025machinelearningassessastrophysical,
      title={Machine Learning to assess astrophysical origin of gravitational waves triggers}, 
      author={Lorenzo Mobilia and Gianluca Maria Guidi},
      year={2025},
      eprint={2509.12882},
      archivePrefix={arXiv},
      primaryClass={gr-qc},
      url={https://arxiv.org/abs/2509.12882}, 
}

@inproceedings{Verma_2022,
   title={Employing deep learning for detection of gravitational waves from compact binary coalescences},
   volume={2609},
   ISSN={0094-243X},
   url={http://dx.doi.org/10.1063/5.0108682},
   DOI={10.1063/5.0108682},
   publisher={AIP Publishing},
   author={Verma, Chetan and Reza, Amit and Krishnaswamy, Dilip and Caudill, Sarah and Gaur, Gurudatt},
   year={2022},
   pages={020010} }

@article{Shawhan_2004,
   title={A new waveform consistency test for gravitational wave inspiral searches},
   volume={21},
   ISSN={1361-6382},
   url={http://dx.doi.org/10.1088/0264-9381/21/20/018},
   DOI={10.1088/0264-9381/21/20/018},
   number={20},
   journal={Classical and Quantum Gravity},
   publisher={IOP Publishing},
   author={Shawhan, Peter and Ochsner, Evan},
   year={2004},
   month=sep, pages={S1757–S1765} }

@misc{patel2025radarradioafterglowdetectionaidriven,
      title={RADAR-Radio Afterglow Detection and AI-driven Response: A Federated Framework for Gravitational Wave Event Follow-Up}, 
      author={Parth Patel and Alessandra Corsi and E. A. Huerta and Kara Merfeld and Victoria Tiki and Zilinghan Li and Tekin Bicer and Kyle Chard and Ryan Chard and Ian T. Foster and Maxime Gonthier and Valerie Hayot-Sasson and Hai Duc Nguyen and Haochen Pan},
      year={2025},
      eprint={2507.14827},
      archivePrefix={arXiv},
      primaryClass={astro-ph.HE},
      url={https://arxiv.org/abs/2507.14827}, 
}

@misc{AttenGWRepo,
  author       = {Tiki, Victoria},
  title         = {{AttenGW}: Public Gravitational-Wave Detection Model and Preprocessing Pipeline},
  note           = {{GitHub} repository},
  year         = {2025},
  url          = {https://github.com/victoria-tiki/AttenGW},
}

@article{Abbott2024GWTC21,
    author        = {Abbott, R. and others},
    collaboration = {LIGO Scientific Collaboration and Virgo Collaboration},
    title         = {{GWTC-2.1: Deep Extended Catalog of Compact Binary
                      Coalescences Observed by LIGO and Virgo During the
                      First Half of the Third Observing Run}},
    journal       = {Phys. Rev. D},
    volume        = {109},
    number        = {2},
    pages         = {022001},
    year          = {2024},
    doi           = {10.1103/PhysRevD.109.022001},
    eprint        = {2108.01045},
    archivePrefix = {arXiv},
    primaryClass  = {gr-qc}
}

@misc{kingma2017adammethodstochasticoptimization,
      title={Adam: A Method for Stochastic Optimization}, 
      author={Diederik P. Kingma and Jimmy Ba},
      year={2017},
      eprint={1412.6980},
      archivePrefix={arXiv},
      primaryClass={cs.LG},
      url={https://arxiv.org/abs/1412.6980}, 
}

\pagebreak
\appendix

\section{Validation of the Original \textsc{AttenGW} Architecture}
\label{app:attengw_validation}

An early version of the cross-attention \textsc{AttenGW} architecture was introduced as the gravitational-wave component of the RADAR federated GW--radio follow-up framework~\cite{patel2025radarradioafterglowdetectionaidriven}. In that work, the detector was demonstrated on a single hour-long segment of LIGO data but was not systematically evaluated over an extended real-noise background. We therefore perform a separate validation of the original architecture using a February 2020 O3b benchmark established in Ref.~\cite{Tian:2023vdc}.

This experiment uses the legacy preprocessing and benchmark configuration associated with that work and is reported separately from the O3a architecture ablations in Sec.~\ref{sec:ablationstudies}. Its purpose is to provide an extended real-noise validation of the original \textsc{AttenGW} model rather than to select the architecture used in the final analysis. The comparison with the graph-based benchmark additionally provides context for evaluating whether the cross-attention implementation improved upon the architecture it was designed to replace.

\subsection{Methods}

\begin{figure}[h!]
    \centering
    \includegraphics[width=0.95\linewidth]{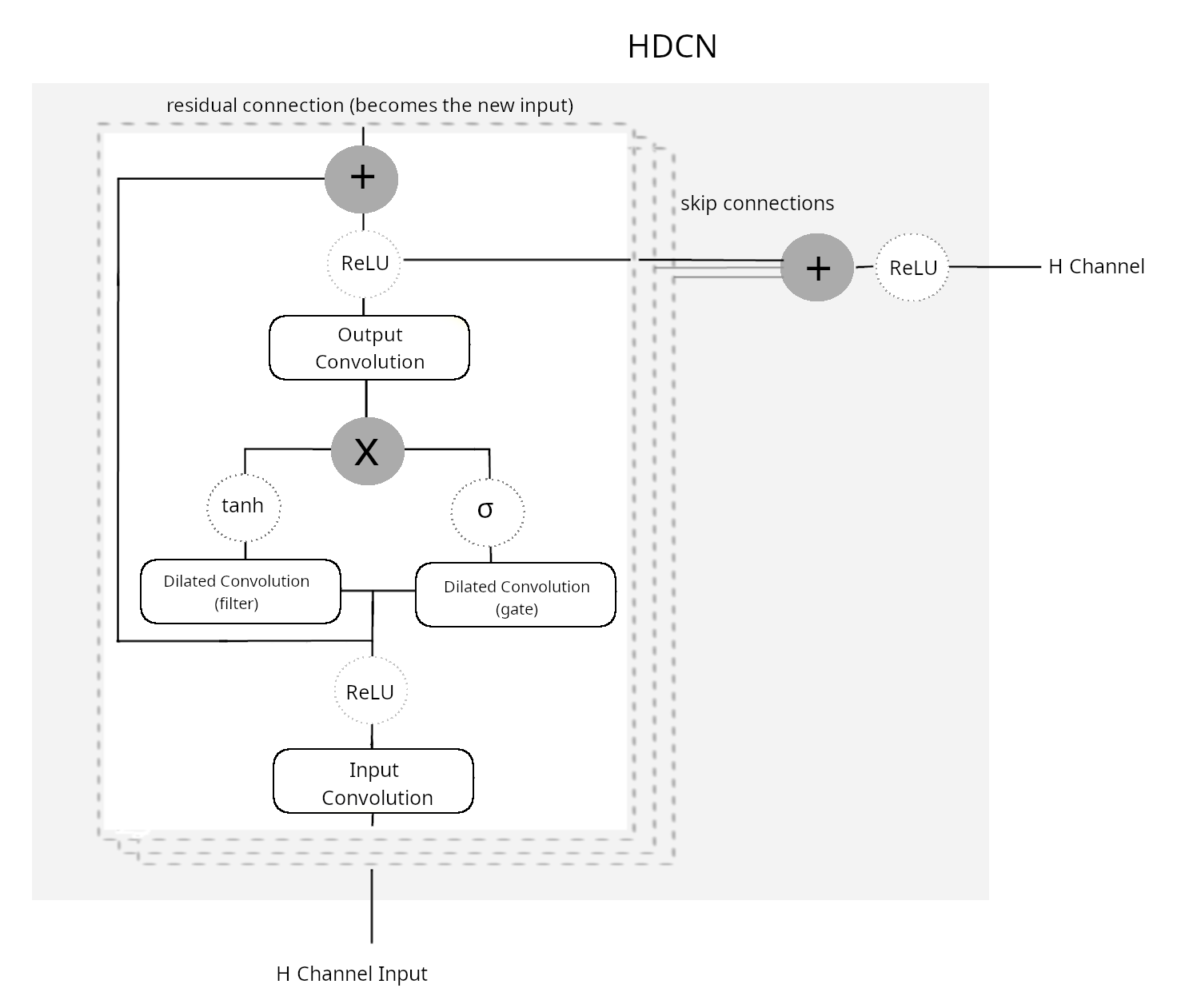}
    \caption{HDCN block applied to the Hanford channel. Each layer applies the same processing steps, except in the first layer, where the initial convolution lifts the one-dimensional input time series to 16 channels.}
    \label{fig:model-schematic_hdcn}
\end{figure}

\begin{figure}[h!]
    \centering
    \includegraphics[width=0.95\linewidth]{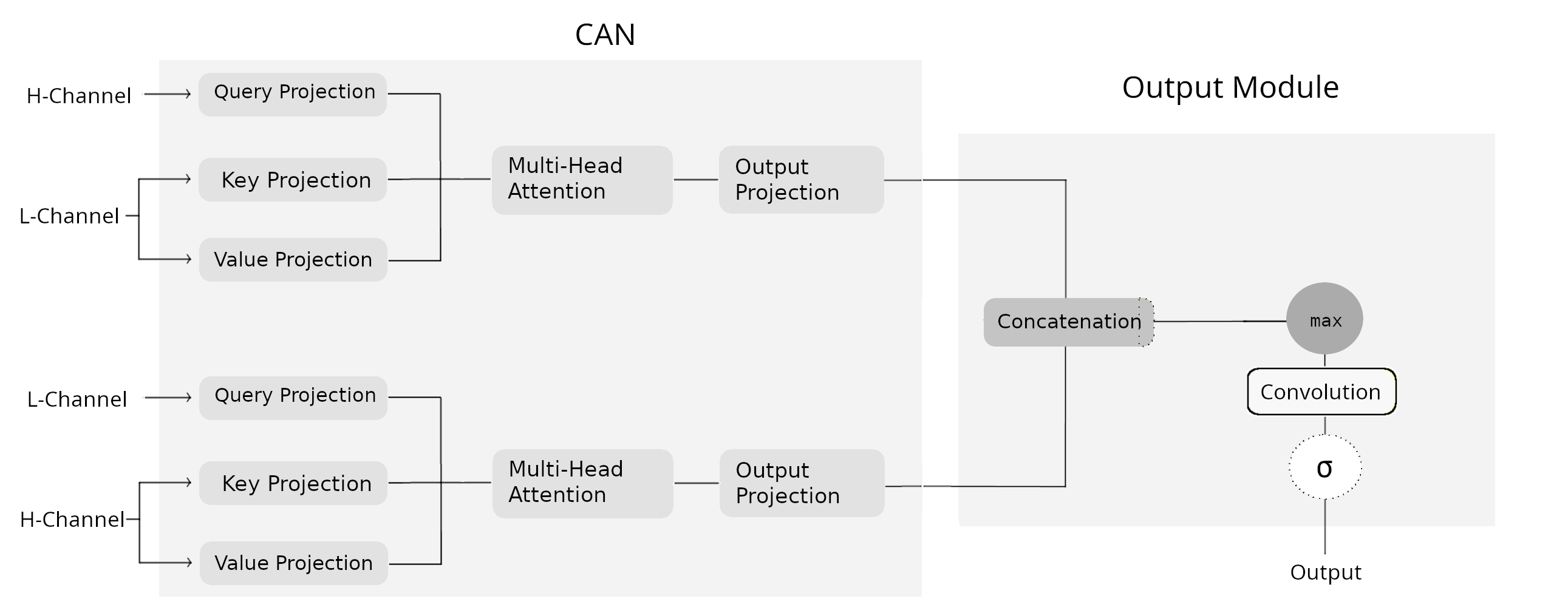}
    \caption{Cross-attention and output modules used to aggregate the detector representations and produce per-timestep detection scores.}
    \label{fig:model-schematic-can}
\end{figure}

The original \textsc{AttenGW} model uses separate WaveNet-style HDCN encoders for Hanford and Livingston, followed by bidirectional cross-detector attention and a per-timestep output head. The resulting 193,000-parameter model produces a scalar detection score at each timestep; its encoder and aggregation modules are shown in Figs.~\ref{fig:model-schematic_hdcn} and~\ref{fig:model-schematic-can}.

For the benchmark comparison, we use the pre-whitened Hanford and Livingston training data and February 2020 inference data supplied by Tian et al., ensuring consistency with their preprocessing. The training data consist of three fixed O3a noise segments beginning at GPS times 1240989696, 1241128960, and 1242738688. Synthetic waveforms are whitened using the PSDs associated with the selected pre-whitened noise file and normalized to unit peak amplitude, while the noise is rescaled to a curriculum-selected time-domain standard deviation before injection; this schedule therefore controls the relative signal and noise amplitudes rather than a matched-filter SNR. Inputs contain $1\,\mathrm{s}$ of strain, and the positive target extends over the $500\,\mathrm{ms}$ preceding the merger. Noise-only examples are drawn with probability 0.65. We train with Adam using an initial learning rate of $10^{-3}$, a weight decay of $10^{-5}$, and a batch size of 32. Inference uses 4096-sample windows with stride 4096 and offsets \([0,2048]\). Candidate triggers are constructed using the \texttt{full\_peak} definition described in Sec.~\ref{inference}, which most closely resembles the trigger criterion used by Tian et al.

Multiple instances of the cross-attention model are trained independently so that we can evaluate both their individual background counts and coincidence among models over the February 2020 evaluation data. The \textsc{AttenGW} implementation uses the two LIGO detectors, whereas the graph-based benchmark architecture used in the original Tian et al. benchmark also incorporates Virgo; differences in detector inputs may therefore contribute to the observed performance and prevent attributing the full difference solely to the aggregation mechanism. Unlike Tian et al., who fine-tuned their models using noise drawn from the same February 2020 period used for the final benchmark, we train exclusively on O3a noise and apply the resulting models unchanged to the O3b evaluation data. We also count every trigger produced by the predefined inference pipeline, without subsequently removing candidates using data-quality information or spectrogram inspection. The experiment therefore tests generalization to a later noise period without adaptation to the evaluation background.

\subsection{Results} 
Individual graph-based models produce background counts on the order of $10^2$--$10^3$ triggers over the February 2020 evaluation data. Under the same benchmark conditions and at the same detection efficiency, a single instance of the cross-attention \textsc{AttenGW} model produces 124 background triggers without reducing the cataloged-signal recovery used in the comparison. Requiring coincidence between two independently trained \textsc{AttenGW} models reduces the background to five triggers. Coincidence across three independently trained models eliminates all observed background triggers while preserving the same detection efficiency. By comparison, Tian et al. used coincidence across six independently trained graph models, after which 13 candidate triggers remained and were reduced to six events through subsequent data-quality and spectrogram-based review. 

Under the February 2020 benchmark, \textsc{AttenGW} thus produced fewer single-model background triggers and reached zero observed background with a smaller ensemble than the graph-based reference, which provides evidence in favor of the cross-attention implementation under those benchmark conditions. The O3a comparison using the new preprocessing and training pipeline developed in this work, shown in Fig.~\ref{fig:additional_seeds}, is less decisive: cross-attention exhibits much greater seed-to-seed variation than the graph model and reaches lower sensitive volume at usable operating points\footnote{However, the HDCN fusion model with temporal self-attention instead of cross-attention used for \textsc{AttenGW} performs more consistently and reaches higher sensitive volume}. The different conclusions likely reflect the extensive differences in preprocessing, underscoring how strongly architectural comparisons depend on the surrounding processing choices. 

Repeating the February 2020 evaluation with the pipeline used in the main analysis substantially improves the graph model: a single trained model produces about 20 false positives with eight recoveries. For comparison, TCN early fusion produces zero false positives with eight or nine recoveries (Sec.~\ref{sec:real_recovery}). Although the graph model remains weaker than TCN early fusion, its substantial improvement under the new pipeline indicates that the earlier behavior was shaped at least as much by the overall pipeline as by the architecture itself. Notably, the current pipeline injects signals into noise before whitening, rather than whitening them separately before injection, which more closely reflects the physical detection process.

Nevertheless, under the revised pipeline, all tested HDCN variants remain weaker than the strongest TCN models, and we therefore do not pursue further HDCN benchmarks.

\section{Effect of Time Shifting on False-Positive Counts}

\begin{figure*}[!h]
    \centering
    \includegraphics[width=0.8\textwidth]
{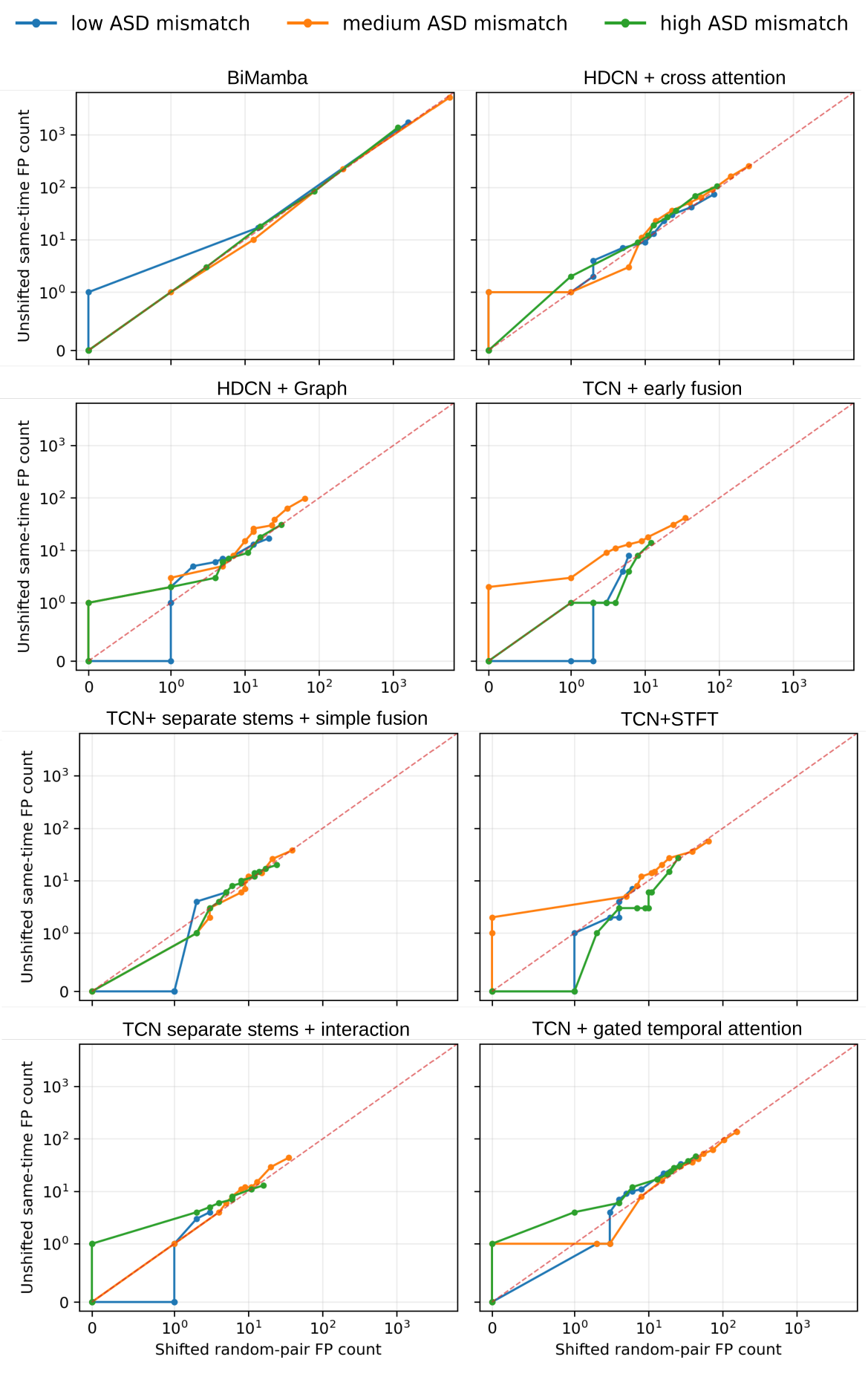}
    \caption{False positives on unshifted versus time-shifted segments across evaluation thresholds for 8 models. }
    \label{fig:shifted_vs_unshifted}
\end{figure*}

In Figure~\ref{fig:shifted_vs_unshifted}, false positives recorded on original, unshifted noise data from August 2019 are compared with those measured on the corresponding time-shifted segments. For various architectures, each curve traces the false positive values obtained as the evaluation threshold is varied. Agreement between the shifted and unshifted results indicates that time shifting does not materially alter false-positive behavior.

Across architectures, false-positive counts from shifted and unshifted data generally follow the one-to-one relation across thresholds. No consistent bias toward higher or lower counts is observed, indicating that time shifting preserves the overall false-positive behavior and supporting its use in constructing the partially synthetic evaluation dataset.

\section{Architecture Comparison and Baseline Selection}
\label{app:additional_seeds}

\begin{figure*}[!h]
    \centering
    \includegraphics[width=\textwidth]{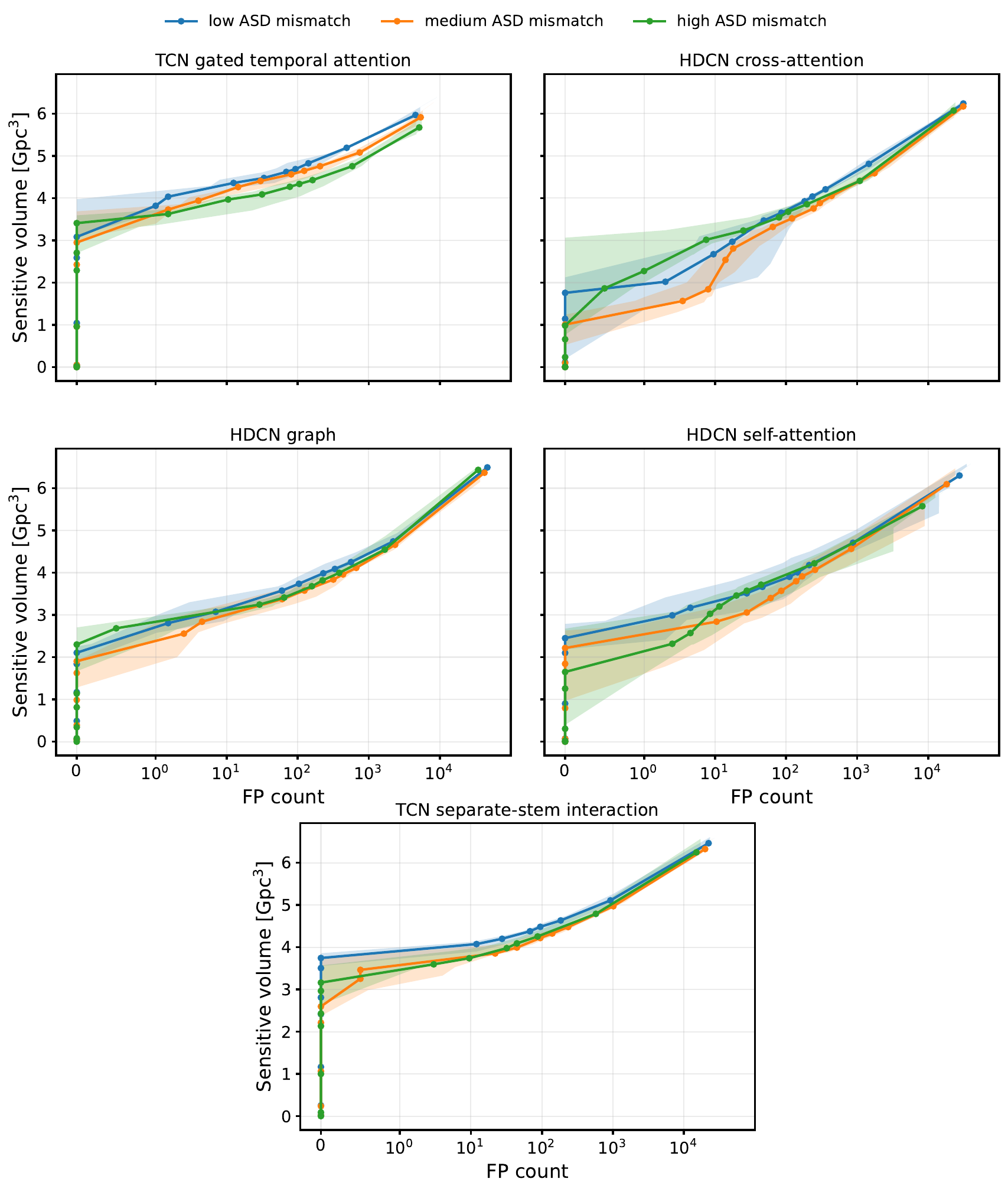}
    \caption{
    Sensitive volume as a function of the number of observed false positives across training seeds for five architectures. Each panel shows the median across seeds for each threshold, with shaded bands indicating the full seed-to-seed range. 
    }
    \label{fig:additional_seeds}
\end{figure*}

To assess the robustness of the architecture comparison to random initialization, we train each model using four independent seeds under the reference configuration described in Sec.~\ref{sec:ablationstudies}, including shared per-window normalization and an 8\,s whitening context. Figure~\ref{fig:additional_seeds} shows the sensitive-volume--versus--false-positive curves separately for the three ASD-mismatch regimes, with each curve representing the median across seeds and the shaded region spanning the full seed-to-seed range.

The three HDCN-based architectures perform substantially worse than all of the TCN-based models, indicating that their greater architectural complexity does not yield improved detection performance in this setting. By contrast, the TCN separate-stem interaction and gated temporal-attention models are competitive with TCN early fusion under the shared-normalization and 8\,s whitening-context configuration used here. Both perform substantially better than in the initial architecture screen, although their relative performance fluctuates modestly across ASD-mismatch bins. Together, these results suggest that performance differences among the strongest TCN variants are small. We therefore retain TCN early fusion as the baseline because it achieves comparable performance with the simplest architecture.

\section{Mass-binned sensitive volume}
\label{app:mass_binned_sensitive_volume}

\begin{figure*}[!h]
    \centering
    \includegraphics[width=\textwidth]{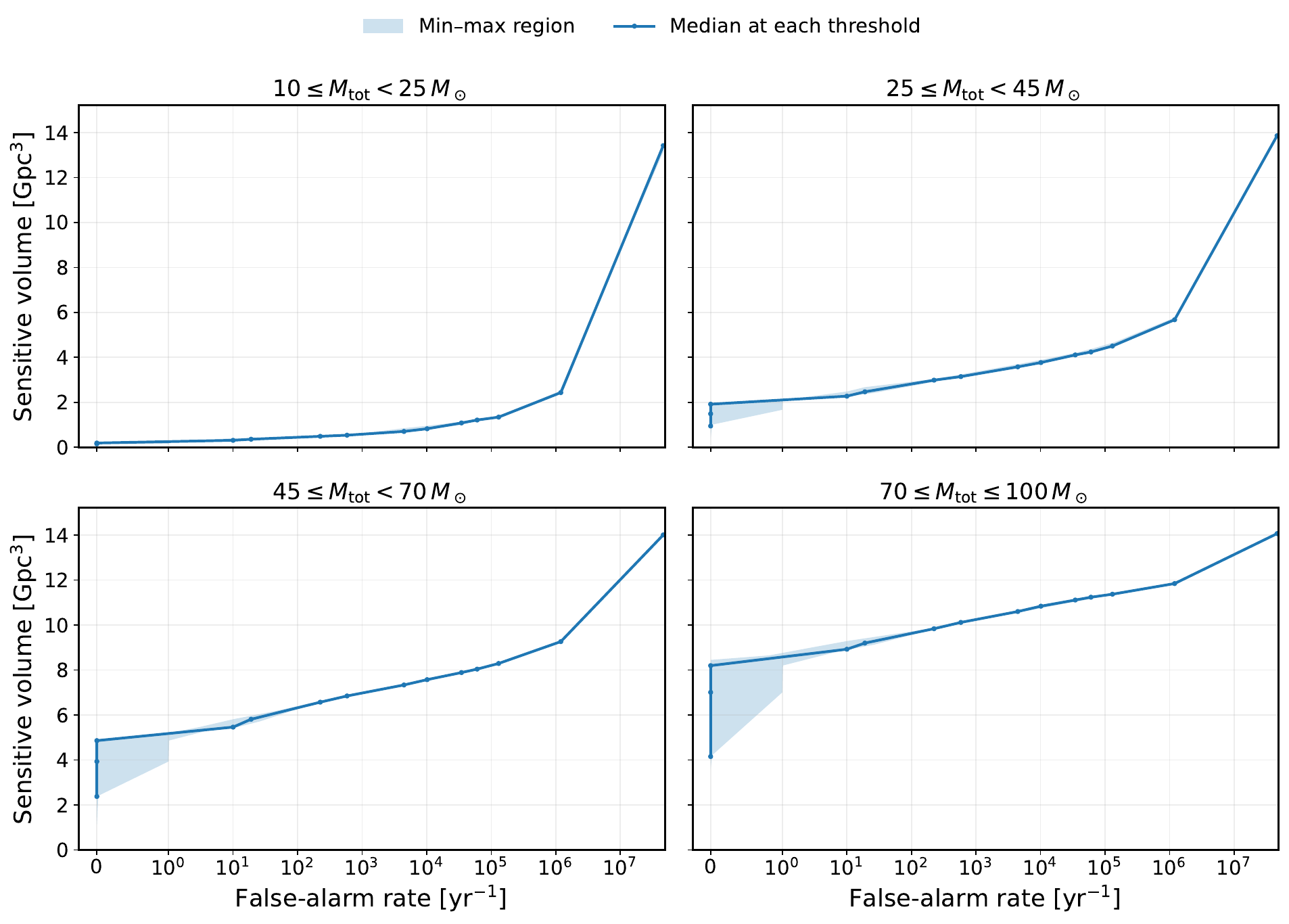}
    \caption{
        Sensitive volume as a function of false-alarm rate in four total-mass bins.
    }
    \label{fig:mass_binned_sensitive_volume}
\end{figure*}

Figure~\ref{fig:mass_binned_sensitive_volume} shows the sensitive-volume results divided into four total-mass bins, using the same false-alarm-rate and seed-summary construction as the main analysis. At each threshold, the solid line shows the median false-alarm rate and median sensitive volume across training seeds, while the shaded region shows the min--max envelope of the interpolated seed curves. As expected, recovery improves with total mass. At the highest false-alarm rates of \(\mathrm{FAR}\sim5\times10^{7}\,\mathrm{yr}^{-1}\), the curves approach the maximum volume \(V_{\mathrm{eff}}\simeq14\,\mathrm{Gpc}^3\) of the evaluated injection distance range because the models trigger in nearly every inference window. This reflects saturation of the sensitive-volume estimator in an operationally irrelevant high-FAR regime, rather than the astrophysical reach of the detector.

\end{document}